\documentclass[aps,prd,preprint,tightenlines,onecolumn,superscriptaddress,amsmath,amssymb]{revtex4}
\usepackage{graphicx} %
\usepackage{epstopdf} %
\usepackage{dcolumn} %
\usepackage{xcolor} %
\usepackage{amsmath,amssymb} %
\usepackage{hyperref} %
\usepackage[utf8]{inputenc} %
\usepackage[english]{babel} %
\usepackage[mathlines]{lineno} %
\usepackage{blindtext} %
\usepackage{ifthen} %
\usepackage{multirow}

\usepackage{orcidlink} %

\usepackage[capitalise]{cleveref} %

\usepackage{afterpage} %
\usepackage{rotating,booktabs} %

\usepackage{setspace}

\newcommand*\patchAmsMathEnvironmentForLineno[1]{%
\expandafter\let\csname old#1\expandafter\endcsname\csname #1\endcsname
\expandafter\let\csname oldend#1\expandafter\endcsname\csname
end#1\endcsname
 \renewenvironment{#1}%
   {\linenomath\csname old#1\endcsname}%
   {\csname oldend#1\endcsname\endlinenomath}%
}
\newcommand*\patchBothAmsMathEnvironmentsForLineno[1]{%
  \patchAmsMathEnvironmentForLineno{#1}%
  \patchAmsMathEnvironmentForLineno{#1*}%
}
\AtBeginDocument{%
\patchBothAmsMathEnvironmentsForLineno{equation}%
\patchBothAmsMathEnvironmentsForLineno{align}%
\patchBothAmsMathEnvironmentsForLineno{flalign}%
\patchBothAmsMathEnvironmentsForLineno{alignat}%
\patchBothAmsMathEnvironmentsForLineno{gather}%
\patchBothAmsMathEnvironmentsForLineno{multline}%
\patchBothAmsMathEnvironmentsForLineno{eqnarray}%
}

\graphicspath{{figures/}} %

\newboolean{articletitles}
\setboolean{articletitles}{true} %

\RequirePackage{xspace}
\RequirePackage{relsize}

\def\belletwo{\mbox{Belle~II}\xspace}

\def\babar{\mbox{\slshape B\kern-0.1em{\smaller A}\kern-0.1em
    B\kern-0.1em{\smaller A\kern-0.2em R}}\xspace}

\def\Pmu         {\ensuremath{\mu}\xspace}                 
\def\Pnu         {\ensuremath{\nu}\xspace}                 
                 
\def\Ppi         {\ensuremath{\pi}\xspace}

\def\Ppsi        {\ensuremath{\psi}\xspace}                 
                 
\mathchardef\PDelta="7101
\mathchardef\PXi="7104
\mathchardef\PLambda="7103
\mathchardef\PSigma="7106
\mathchardef\POmega="710A
\mathchardef\PUpsilon="7107
                 
\def\PB      {\ensuremath{B}\xspace}                 
                 
\def\PD      {\ensuremath{D}\xspace}

\def\PJ      {\ensuremath{J}\xspace}                 
\def\PK      {\ensuremath{K}\xspace}

\def\Pb      {\ensuremath{b}\xspace}                 
\def\Pc      {\ensuremath{c}\xspace}                 
                 
\def\Pe      {\ensuremath{e}\xspace}

\def\Pi      {\ensuremath{i}\xspace}

\def\Pp      {\ensuremath{p}\xspace}

\def\Ps      {\ensuremath{s}\xspace}

\def\en         {\ensuremath{\Pe^-}\xspace}   %

\def\epem       {\ensuremath{\Pe^+\Pe^-}\xspace}

\def\mun        {\ensuremath{\Pmu^-}\xspace} %

\def\neu        {\ensuremath{\Pnu}\xspace}
\def\neub       {\ensuremath{\overline{\Pnu}}\xspace}

\def\squark    {\ensuremath{\Ps}\xspace}
\def\squarkbar {\ensuremath{\overline \squark}\xspace}

\def\cquark    {\ensuremath{\Pc}\xspace}
\def\cquarkbar {\ensuremath{\overline \cquark}\xspace}
\def\ccbar     {\ensuremath{\cquark\cquarkbar}\xspace}
\def\bquark    {\ensuremath{\Pb}\xspace}

\def\pion  {\ensuremath{\Ppi}\xspace}
\def\piz   {\ensuremath{\pion^0}\xspace}

\def\pip   {\ensuremath{\pion^+}\xspace}
\def\pim   {\ensuremath{\pion^-}\xspace}

\def\kaon  {\ensuremath{\PK}\xspace}
\def\Kbar  {\kern 0.2em\overline{\kern -0.2em \PK}{}\xspace}%

\def\Kz    {\ensuremath{\kaon^0}\xspace}
\def\Kzb   {\ensuremath{\Kbar^0}\xspace}
\def\KzKzb {\ensuremath{\Kz \kern -0.16em \Kzb}\xspace}
\def\Kp    {\ensuremath{\kaon^+}\xspace}
\def\Km    {\ensuremath{\kaon^-}\xspace}

\def\KpKm  {\ensuremath{\Kp \kern -0.16em \Km}\xspace}
\def\KS    {\ensuremath{\kaon^0_{\rm\scriptscriptstyle S}}\xspace} 
\def\KL    {\ensuremath{\kaon^0_{\rm\scriptscriptstyle L}}\xspace}

\def\D       {\ensuremath{\PD}\xspace}
\def\Dbar    {\kern 0.2em\overline{\kern -0.2em \PD}{}\xspace}%

\def\Dz      {\ensuremath{\D^0}\xspace}
\def\Dzb     {\ensuremath{\Dbar^0}\xspace}
\def\DzDzb   {\ensuremath{\Dz {\kern -0.16em \Dzb}}\xspace}
\def\Dp      {\ensuremath{\D^+}\xspace}
\def\Dm      {\ensuremath{\D^-}\xspace}

\def\DpDm    {\ensuremath{\Dp {\kern -0.16em \Dm}}\xspace}

\def\Dstarp  {\ensuremath{\D^{*+}}\xspace}

\def\Dstarpm {\ensuremath{\D^{*\pm}}\xspace}

\def\Dsp     {\ensuremath{\D^+_\squark}\xspace}

\def\B       {\ensuremath{\PB}\xspace}
\def\Bbar    {\ensuremath{\kern 0.18em\overline{\kern -0.18em \PB}{}}\xspace}%

\def\Bz      {\ensuremath{\B^0}\xspace}
\def\Bzb     {\ensuremath{\Bbar^0}\xspace}

\def\Bs      {\ensuremath{\B^0_\squark}\xspace}
\def\Bsb     {\ensuremath{\Bbar^0_\squark}\xspace}

\def\jpsi     {\ensuremath{{\PJ\mskip -3mu/\mskip -2mu\Ppsi\mskip 2mu}}\xspace}

\def\Y#1S{\ensuremath{\PUpsilon{(#1S)}}\xspace}%

\def\proton      {\ensuremath{\Pp}\xspace}
\def\antiproton  {\ensuremath{\overline \proton}\xspace}

\def\Xires {\ensuremath{\PXi}\xspace}

\def\L {\ensuremath{\PLambda}\xspace}
\def\Lbar {\ensuremath{\kern 0.1em\overline{\kern -0.1em\PLambda}}\xspace}

\def\Omegares {\ensuremath{\POmega}\xspace}

\def\Lc      {\ensuremath{\L^+_\cquark}\xspace}

\def\to                 {\ensuremath{\rightarrow}\xspace}

\def\order   {\ensuremath{\mathcal{O}}\xspace}

\def\CP                {\ensuremath{C\!P}\xspace}

\newcommand{\efftag}{{\ensuremath{\varepsilon_{\rm tag}}}\xspace}
\newcommand{\mistag}{\ensuremath{\omega}\xspace}

\newcommand{\efftageff}{{\ensuremath{\varepsilon^{\rm eff}_{\rm tag}}}\xspace}

\newcommand{\tev}{\ensuremath{\mathrm{\,Te\kern -0.1em V}}\xspace}
\newcommand{\gev}{\ensuremath{\mathrm{\,Ge\kern -0.1em V}}\xspace}
\newcommand{\mev}{\ensuremath{\mathrm{\,Me\kern -0.1em V}}\xspace}
\newcommand{\kev}{\ensuremath{\mathrm{\,ke\kern -0.1em V}}\xspace}
\newcommand{\ev}{\ensuremath{\mathrm{\,e\kern -0.1em V}}\xspace}
\newcommand{\gevc}{\ensuremath{{\mathrm{\,Ge\kern -0.1em V\!/}c}}\xspace}
\newcommand{\mevc}{\ensuremath{{\mathrm{\,Me\kern -0.1em V\!/}c}}\xspace}
\newcommand{\gevcc}{\ensuremath{{\mathrm{\,Ge\kern -0.1em V\!/}c^2}}\xspace}
\newcommand{\gevgevcccc}{\ensuremath{{\mathrm{\,Ge\kern -0.1em V^2\!/}c^4}}\xspace}
\newcommand{\mevcc}{\ensuremath{{\mathrm{\,Me\kern -0.1em V\!/}c^2}}\xspace}

\def\cm   {\ensuremath{\rm \,cm}\xspace}

\def\invfb   {\ensuremath{\mbox{\,fb}^{-1}}\xspace}

\newcommand{\stat}{\ensuremath{\mathrm{(stat)}}\xspace}
\newcommand{\syst}{\ensuremath{\mathrm{(syst)}}\xspace}

\def\order{{\ensuremath{\cal O}}\xspace}

\def\gsim{{~\raise.15em\hbox{$>$}\kern-.85em
          \lower.35em\hbox{$\sim$}~}\xspace}
\def\lsim{{~\raise.15em\hbox{$<$}\kern-.85em
          \lower.35em\hbox{$\sim$}~}\xspace}

\newcommand{\mean}[1]{\ensuremath{\left\langle #1 \right\rangle}} %

\def\sPlot{\mbox{\em sPlot}\xspace}

\newcommand{\eg}{\mbox{\itshape e.g.}\xspace}
\newcommand{\ie}{\mbox{\itshape i.e.}\xspace}

\newcommand{\qtrue}{\ensuremath{q_{\rm true}}\xspace}
\newcommand{\defftag}{\ensuremath{\Delta\efftag}\xspace}
\newcommand{\dmistag}{\ensuremath{\Delta\mistag}\xspace}

\newcommand{\pis}{\ensuremath{\pi_{\rm s}}\xspace}
\newcommand{\pbar}{\antiproton}

\newcommand{\lumi}{362\invfb}

\newcommand{\result}{\ensuremath{(47.91\pm 0.07\stat\pm 0.51\syst)\%}\xspace} 
 
\renewcommand{\toprule}{\hline\hline}
\renewcommand{\midrule}{\hline}
\renewcommand{\bottomrule}{\hline\hline}

\makeatletter%
\begin{document}

\includegraphics[width=3cm]{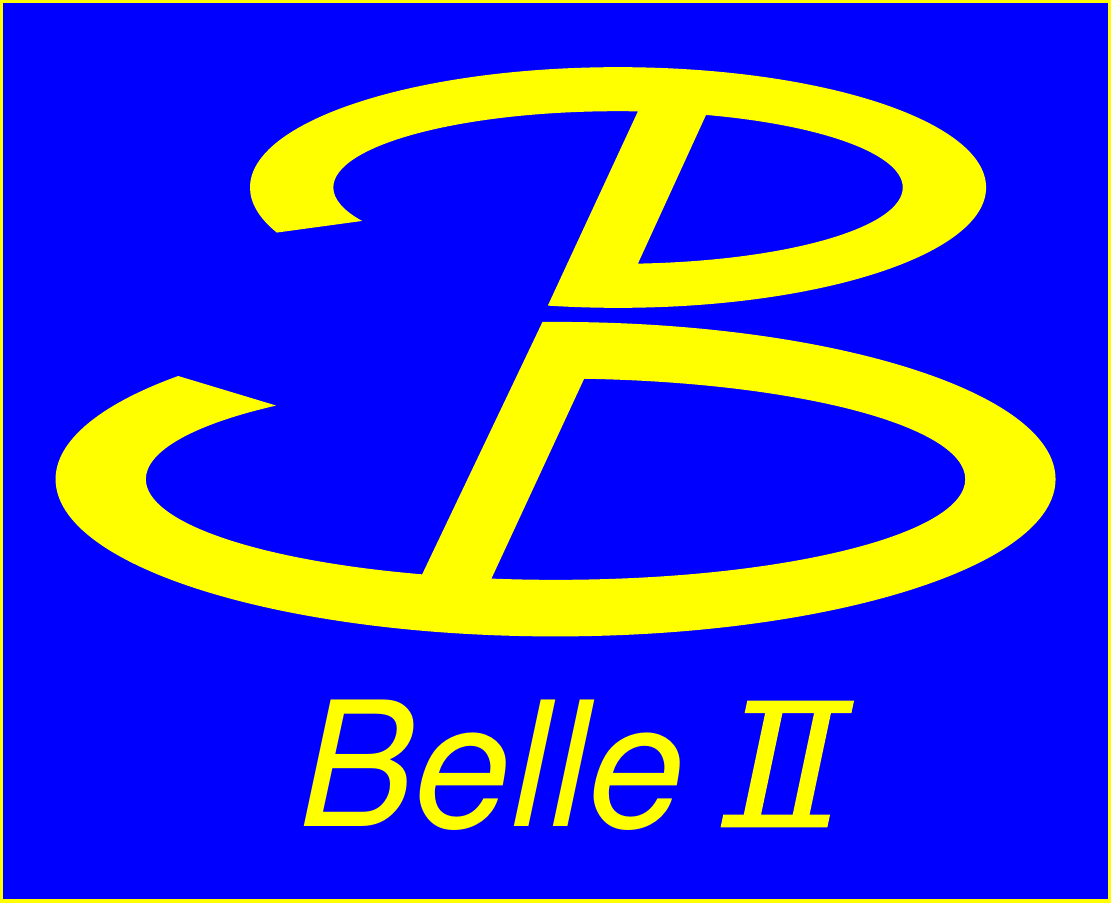}\vspace*{-1.9cm}

\begin{flushright}
KEK preprint: 2023-1\\
Belle II preprint: 2023-005\\
\end{flushright}
\vspace*{32pt}

\title{%
%
Novel method for the identification of the production flavor of neutral charmed mesons 
}
%
%
%
%
%
%
%
  \author{I.~Adachi\,\orcidlink{0000-0003-2287-0173}} %
  \author{L.~Aggarwal\,\orcidlink{0000-0002-0909-7537}} %
  \author{H.~Ahmed\,\orcidlink{0000-0003-3976-7498}} %
  \author{H.~Aihara\,\orcidlink{0000-0002-1907-5964}} %
  \author{N.~Akopov\,\orcidlink{0000-0002-4425-2096}} %
  \author{A.~Aloisio\,\orcidlink{0000-0002-3883-6693}} %
  \author{N.~Anh~Ky\,\orcidlink{0000-0003-0471-197X}} %
  \author{D.~M.~Asner\,\orcidlink{0000-0002-1586-5790}} %
  \author{H.~Atmacan\,\orcidlink{0000-0003-2435-501X}} %
  \author{T.~Aushev\,\orcidlink{0000-0002-6347-7055}} %
% ...
%
\collaboration{The Belle II Collaboration}
 \begin{abstract}
%
%
\noindent We propose a new algorithm for the identification of the production flavor of neutral \D mesons in the \belletwo experiment. The algorithm exploits the correlation between the flavor of a reconstructed neutral \D meson (signal \D meson) and the electric charges of particles reconstructed in the rest of the $\epem\to\ccbar$ event. These include those originating from the decay of the other charm hadron produced in the event, as well as those possibly produced in association with the signal \D meson. We develop the algorithm using simulation and calibrate it in data using decay modes that identify the flavor of the decaying neutral \D meson. We use a data sample of \epem collisions, corresponding to \lumi of integrated luminosity, collected by \belletwo at center-of-mass energies near the $\Upsilon(4S)$ mass. The effective tagging efficiency in data is \result, independent of the neutral-\D-meson decay mode. This charm flavor tagger will approximately double the effective sample size of many \CP-violation and charm-mixing measurements that so far have exclusively relied on neutral \D mesons originating from \Dstarpm decays. While developed for \belletwo, the basic principles underlying the charm flavor tagger can be used in other experiments, including those at hadron colliders. 
 \end{abstract}

\maketitle

%
\section{Introduction}%
The violation of charge-conjugation and parity (\CP) symmetry in the up-type quark sector is strongly suppressed in the standard model. Hence weak decays of charmed hadrons offer an excellent opportunity for model-independent searches for non-standard-model physics that are complementary to searches performed in down-type quark transitions using kaons or bottom hadrons~\cite{Grossman:2006jg}. The recent observation of \CP violation in neutral \D-meson decays~\cite{LHCb:2019hro} has stimulated a debate on whether the observed value is consistent with the standard model or not~\cite{Chala:2019fdb,Grossman:2019xcj,Buccella:2019kpn,Li:2019hho,Schacht:2021jaz,Cheng:2019ggx,Dery:2019ysp,Bause:2020obd,Dery:2021mll,Cheng:2021yrn,Bediaga:2022sxw}. Further precise measurements of \CP asymmetries in other charmed-hadron decay modes are needed to clarify the picture, as are searches for yet-to-be-observed signs of \CP violation in \Dz-\Dzb mixing. Measuring mixing and \CP asymmetries in decays of neutral \D mesons typically requires the identification of the charm flavor at production, \ie, whether the neutral \D meson is produced as a \Dz or a \Dzb. This task, known as \textit{flavor tagging}, is accomplished by selecting \Dz mesons that either originate from the strong-interaction decay $\Dstarp\to\Dz\pip$ or from the semileptonic decay of a $b$ hadron $H_b\to \Dz\ell^-\nu X$, where $X$ indicates any other final-state particles that may or may not be reconstructed and $\ell$ is an electron or muon. (Charge-conjugate processes are implied throughout, unless specified otherwise.) Hence, the sample of neutral \D mesons available for measurements that require tagging is much smaller than the inclusive sample of neutral \D mesons produced in \epem or hadron collisions. For example, a typical analysis in \epem collisions at $\sqrt{s}\approx10\gev$ reconstructs approximately five times fewer $\Dstarp\to\Dz(\to\Km\pip)\pip$ decays than untagged $\Dz\to\Km\pip$ decays~\cite{Belle:2008ddg}. The ratio between \Dstarp-tagged and untagged \Dz decays typically reconstructed in hadron collisions is even smaller~\cite{CDF:2011ejf,LHCb:2015swx}.

\begin{figure*}[ht]
\centering
\includegraphics[width=0.7\textwidth]{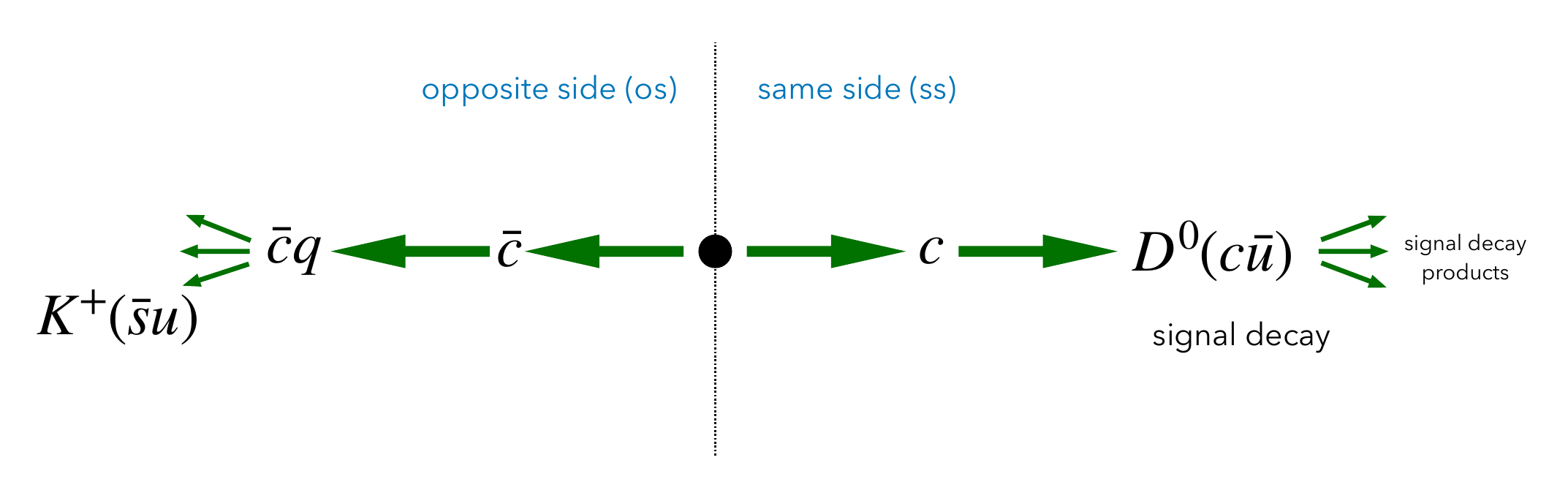}\\
\caption{Simplified representation of an $\epem\to \cquark\cquarkbar$ process in the center-of-mass frame. The charged kaon from the Cabibbo-favored $\cquarkbar\to\squarkbar$ transition on the opposite side tags the production flavor of the signal neutral \D meson.}\label{fig:tagger_princip}
\end{figure*}

In this paper we describe a novel approach to charm-flavor tagging developed for the \belletwo experiment~\cite{Abe:2010gxa}. The approach exploits information from the other charmed hadron produced in the $\epem\to\ccbar$ event (\textit{opposite-side} tagging). 
This information adds to that provided by the previously used \textit{same-side} tagging, which is provided by the charge of the low-momentum (\textit{soft}) pion originating from the decay of a parent \Dstarp meson. Such opposite-side tagging has long been used for \bquark-flavor tagging at both \epem and hadron colliders~\cite{Workman:2022ynf}, but has never been applied to charm-flavor tagging. It relies on the pair production of \cquark and \cquarkbar quarks and infers the flavor of a given neutral \D meson (\textit{signal} \D meson) from the identification of the flavor of the other charmed hadron. The flavor of the other charmed hadron is inferred using the charge of its decay products, which are either charged hadrons ($\pi$, $K$ and $p$) or leptons ($e$ and $\mu$). One example process is schematically shown in \cref{fig:tagger_princip}. The signal \Dz meson arises from the hadronization of a charm quark. The anticharm quark hadronizes into an anticharmed meson (or baryon), which subsequently decays via the Cabibbo-favored $\cquarkbar\to\squarkbar$ transition. As a result, a positively charged kaon is produced. The positive charge of this kaon implies that the signal meson is produced as a \Dz rather than as a \Dzb.

The charm flavor tagger (CFT) presented here infers the flavor of a given signal \D candidate at production by using information from charged particles not associated with the signal decay. These are part of the rest of the event (ROE) and include both opposite-side and same-side particles. As a result, the CFT incorporates the conventional \Dstarp tagging method. We use charged particles in the ROE that are likely to be correlated with the signal flavor to build a set of discriminating variables, which are then input into a binary classifier that provides the tagging decision. The classifier is trained using simulation. The probability of the tagging decision being incorrect, for example due to the possibility of a flavor oscillation or a suppressed decay of the opposite-side charmed hadron, is predicted by the algorithm along with the tagging decision. 

The paper is organized as follows. \Cref{sec:detector} describes the \belletwo detector and simulation. \Cref{sec:tagger} discusses the details of the tagging algorithm, which is trained using simulation. In \cref{sec:performance} we evaluate its performance using several decay modes of charmed hadrons reconstructed in data. For this purpose we use decay modes that identify the flavor of the decaying hadron from their final state particles (\textit{self-tagging} decays). The potential impact of the CFT on physics analyses is estimated in \cref{sec:impact}, followed by the conclusions.
 %
\section{Detector and simulation\label{sec:detector}}
The \belletwo detector~\cite{Abe:2010gxa} surrounds the collision point of the SuperKEKB asymmetric-energy \epem collider~\cite{Akai:2018mbz} and consists of subsystems arranged in a cylindrical geometry around the beam pipe. The innermost is a tracking subsystem consisting of a two-layer silicon-pixel detector surrounded by a four-layer double-sided silicon-strip detector and a 56-layer central drift chamber. Only 15\% of the azimuthal angle is covered by the second pixel detector layer for the data used in this paper. A time-of-propagation counter in the barrel and an aerogel ring-imaging Cherenkov detector in the forward end cap provide information used for the identification of charged particles. An electromagnetic calorimeter consisting of CsI(Tl) crystals fills the remaining volume inside a $1.5\,\rm{T}$ superconducting solenoid and provides energy and timing measurements for photons and electrons. A \KL and muon detection subsystem is installed in the iron flux return of the solenoid. The $z$ axis of the laboratory frame is defined as the central axis of the solenoid, with its positive direction defined as the direction opposite the positron beam.

Simulation is used to train the binary classifier that provides the tagging prediction. The simulation uses \textsc{KKMC}~\cite{Jadach:1999vf} to generate quark-antiquark pairs from \epem collisions, \textsc{Pythia8}~\cite{Sjostrand:2014zea} to simulate the quark hadronization, \textsc{EvtGen}~\cite{Lange:2001uf} to decay the hadrons, and \textsc{GEANT4}~\cite{Agostinelli:2002hh} to simulate the detector response. Events are reconstructed using the \belletwo software~\cite{Kuhr:2018lps,basf2-zenodo}.
 %
\section{The tagging algorithm\label{sec:tagger}}
We introduce the standard metrics used for the evaluation of the tagging performance: the tagging efficiency \efftag and the mistag fraction \mistag. They are defined from the numbers of correctly tagged ($R$, right), wrongly tagged ($W$), and untagged ($U$) candidates in a sample as
\begin{equation}\label{eq:efftag}
\efftag = \frac{R+W}{R+W+U}
\end{equation}
and
\begin{equation}\label{eq:mistag}
\mistag=\left\{\begin{array}{ll}
\dfrac{W}{R+W} &\text{if } W\leqslant R \\
1-\dfrac{W}{R+W} & \text{otherwise}
\end{array}\right.\,.
\end{equation}
\Cref{eq:mistag} implies that $\omega$ cannot exceed 50\% because, whenever $W>R$, the tagging decision is reversed. The tagging performance is quantified by the effective tagging efficiency, or tagging power,
\begin{equation}\label{eq:tagpower}
\efftageff = \efftag(1-2\mistag)^2 = \efftag \mean{r}^2\,,
\end{equation}
where
\begin{equation}\label{eq:dilution}
\mean{r} = 1-2\mistag
\end{equation}
is an average dilution factor that accounts for candidates that are mistagged (see, \eg, \namecref{sec:detector} 8 of Ref.~\cite{Bevan:2014iga}). This nomenclature, which is the standard for flavor tagging algorithms, has the counterintuitive consequence that a small dilution factor has a larger impact on the performance than a large dilution factor. Indeed, a dilution factor $r=0$ indicates that it is not possible to identify the flavor (\ie, the tagging decision is equivalent to random guessing), while a dilution factor $r=1$ indicates that the flavor is perfectly known. The tagging power represents the effective sample size when a tagging decision is required. Typical values of \efftageff for $b$-flavor-tagging algorithms are 20\%--30\% at $e^+e^-$ colliders~\cite{ALEPH:1998kod,DELPHI:2000gjz,SLD:2002dzm,BaBar:2004god,Belle:2005lnu,Belle-II:2021zvj} and 2\%--10\% at hadron colliders~\cite{D0:2008axd,CDF:2011af,CDF:2012nqr,LHCb:2014ini,LHCb:2016inx,LHCb:2019nin,ATLAS:2020lbz,CMS:2020efq}.

The output of the CFT algorithm presented here is a prediction of the product $qr$ of the tagging decision $q$ and the per-candidate dilution factor $r$. We define the tagging decision to be $q=+1$ for signal \Dz decays and $q=-1$ for signal \Dzb decays. The algorithm consists of two steps: (1) ROE charged particles likely to be correlated with the signal flavor are reconstructed and ranked, (2) a binary classifier predicts the product $qr$ from a set of discriminating variables related to the selected ROE particles. The details of the two steps are discussed in the following.

\subsection{Reconstruction and ranking of tagging particles}
For a given signal \D candidate, ROE charged particles are selected by requiring their distances of closest approach to the interaction point to be smaller than $1\cm$ in the transverse plane and smaller than $3\cm$ in the longitudinal direction. Simulation shows that such requirements select on average six ROE charged particles per event in $\epem\to\ccbar$ events at \belletwo. The ROE particles are classified into two groups depending on their electric charge and ranked according to their opening angle with respect to the momentum of the signal \Dz meson in the \epem center-of-mass frame. Momenta of particles emerging from the decay of the other charmed hadron or from the decay of a parent \Dstarp meson are expected to be nearly collinear with the momentum of the signal charmed meson and are highest ranked. We retain only the three top-ranked positively charged particles and the three top-ranked negatively charged particles for subsequent analysis. Keeping more particles does not improve the performance, while keeping fewer reduces it. We label them as $1^+$, $2^+$, $3^+$ and $1^-$, $2^-$, $3^-$, respectively. If an event contains fewer than three positively or negatively charged ROE particles, the associated input variables for the missing particles are represented as missing values. Events that do not contain any ROE particle are not tagged by the CFT, \ie, they reduce the tagging efficiency.

\subsection{Tagging prediction}
The CFT uses a binary classification algorithm to predict the product $qr$. Its input variables are a set of reconstructed quantities from the selected set of six top-ranked ROE particles. In the training of the algorithm, we use generator-level information to label the input ROE particles according to the relations between their electric charges and the signal flavor. The labeling process starts by categorizing the ROE particles according to their species, to whether they originated from decays of charmed hadrons or not, and from which charmed hadron they arose. We consider the categories shown in \cref{tab:tagging-categories} and use the signal \Dz meson to train the algorithm if at least one of the six top-ranked ROE particles falls in one of the categories. This requirement typically removes 10\% of the available $\epem\to\ccbar$ events. The categories are used only to determine which events are used for training. The algorithm does not need to associate the ROE particles to a category to estimate $qr$. For this reason, charged particles produced in the fragmentation of the \ccbar pair, although not used in the training, contribute to the tagging decision.

\begin{table}[t]
\centering
\begin{tabular}{ccccc}
\toprule
Particle & Parent or & \multicolumn{3}{c}{Ranking} \\
type &  grandparent & $1$ & $2$ & $3$ \\
\midrule
\Km & \Dz, \Dp, \Dsp, \Lc & 8.2\% & 8.2\% & 3.6\% \\
\mun, \en & \Dz, \Dp, \Dsp, \Lc & 4.1\% & 3.0\% & 1.4\%  \\
$p$ & \Lc, $\Xires_c^+$, $\Xires_c^0$, $\Omegares_c^0$ & 0.7\% & 0.5\% & 0.2\% \\
$\pis^+$ & \Dstarp (only parent) & 15.4\% & 3.5\% & 0.9\% \\
$\pi^+$ & \Dz, \Dp, \Dsp, \Lc & 22.8\% & 21.0\% & 11.9\% \\
\bottomrule
\end{tabular}
\caption{Tagging categories used for labeling the ROE particles used in the training of the CFT. In each category the ROE particle is requested to match the given species and to originate from the decay of the given parent or grandparent particle. The symbol \pis is used to indicate the soft pion from $\Dstarp\to\Dz\pip$ decays. The relative occurrence at which the $i$-th ($i=1,2,3$) top-ranked ROE particle falls in a category is given in the last three columns (each column sums to 100\% with the inclusion of the remaining cases in which the $i$-th top-ranked ROE particle does not fall into any category).\label{tab:tagging-categories}}
\end{table}

\begin{table}[t]
\centering
\begin{tabular}{cccc}
\toprule
\qtrue & Signal flavor &  \multicolumn{2}{c}{ROE particle charge} \\
& & \multicolumn{1}{c}{Same side} & Opposite side \\
\midrule
$+1$ & \Dz & $\pis^+$ & $K^+,\mu^-,e^-,\pis^-,\pbar,\pi^-$ \\ 
$-1$ & \Dzb & $\pis^-$ & $K^-,\mu^+,e^+,\pis^+,p,\pi^+$ \\
\bottomrule
\end{tabular}
\caption{Relations between the true tagging label \qtrue, the signal flavor, and the charge of the ROE particle.}\label{tab:flavor-charge-relations}
\end{table}

Using basic physics principles relating the ROE particles to the decay of charmed hadrons, the true tagging label \qtrue (either $+1$ or $-1$ for \Dz and \Dzb, respectively) is assigned using the flavor-charge relations of \cref{tab:flavor-charge-relations}. For soft pions, the relations depend on whether the ROE particle is on the same or opposite side as the signal. A soft-pion in the ROE is considered to be on the same side as the signal if its parent \Dstarp meson is also the parent of the signal \Dz meson. 

As a classification algorithm, we use a histogram-based gradient-boosting decision tree (HBDT) from the \texttt{scikit-learn} library~\cite{scikit-learn}, which is particularly suited for large samples as every input variable is binned before training. In addition, this algorithm supports missing input values making it straightforward to include events with fewer than six ROE particles. We configure the HBDT by setting the maximum depth of each tree to 10, the maximum number of leaves for each tree to 31, the minimum number of samples per leaf to 20, the maximum number of bins to use for non-missing values to 255 and the learning rate to 0.2. These values optimize performance while avoiding overtraining.

For every ROE particle, the HBDT receives two input variables: the angle $\Delta R =\sqrt{(\Delta \phi)^2+(\Delta \eta)^2}$, where $\Delta \phi$ denotes the difference in azimuthal angles and $\Delta \eta$ the difference in pseudorapidities between the momentum of the ROE particle and that of the signal \Dz meson, and the difference between the pion and kaon identification discriminators, $P_\pi-P_K$. The particle identification relies on information from all subdetector systems to construct likelihoods $\mathcal{L}(x)$ for a given particle hypothesis $x=e,\mu,\pi,K,p,d$. The particle identification discriminator is defined as the ratio between the likelihood in a given hypothesis and the sum of the likelihoods in all hypotheses, $P_x=\mathcal{L}(x)/\sum_y\mathcal{L}(y)$. For the highest-ranked positively and negatively charged ROE particles we use in addition the invariant mass of the system recoiling against the ROE particle, $m_{\rm recoil}=\sqrt{(p_{e^+e^-}-p_\text{ROE particle})^2}$ with $p$ indicating the four-momentum (computed using the pion mass hypothesis), to make up a total of 14 input variables.

\begin{figure*}[hp]
\centering
\includegraphics[width=0.33\textwidth]{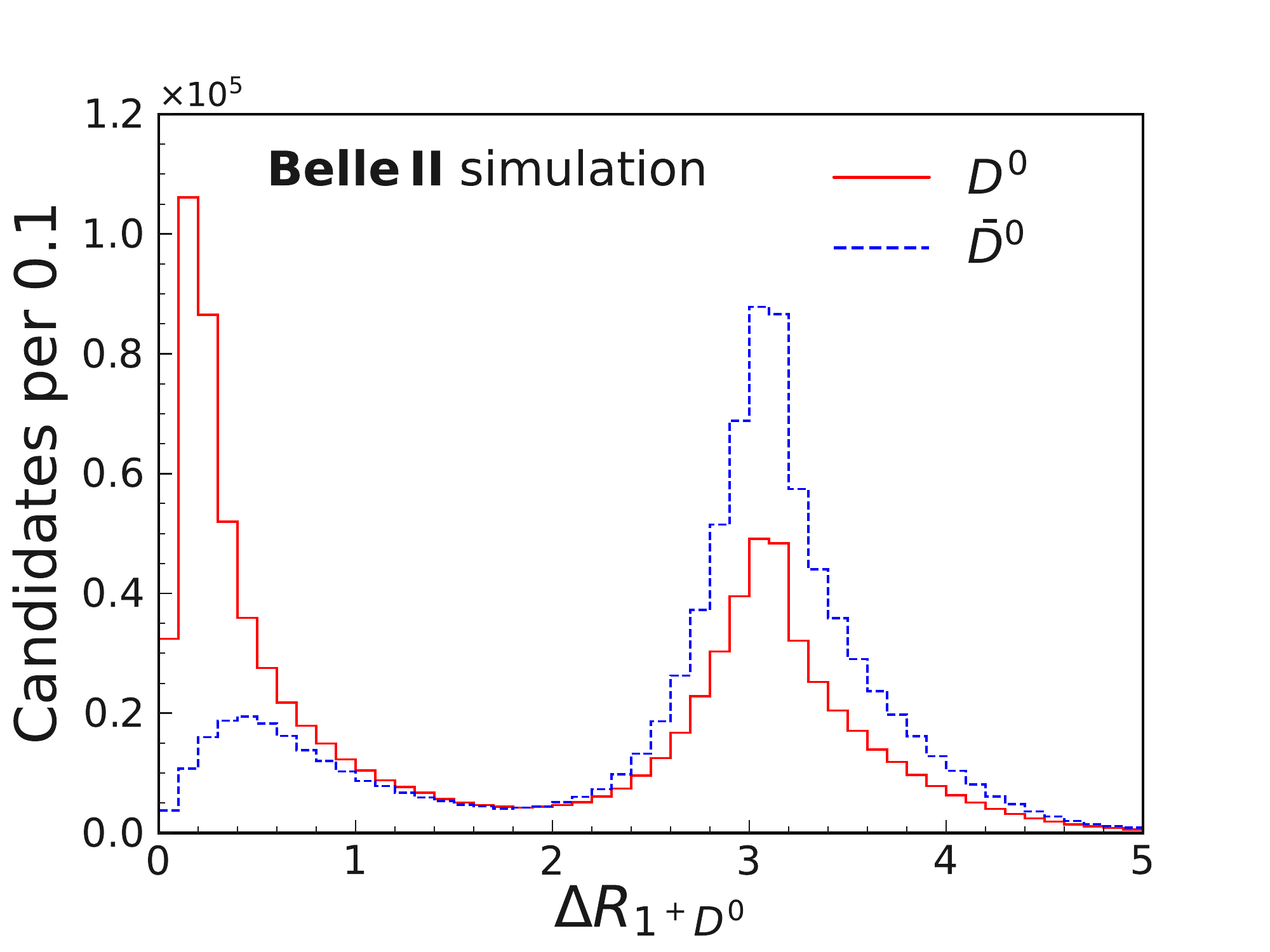}\hfil
\includegraphics[width=0.33\textwidth]{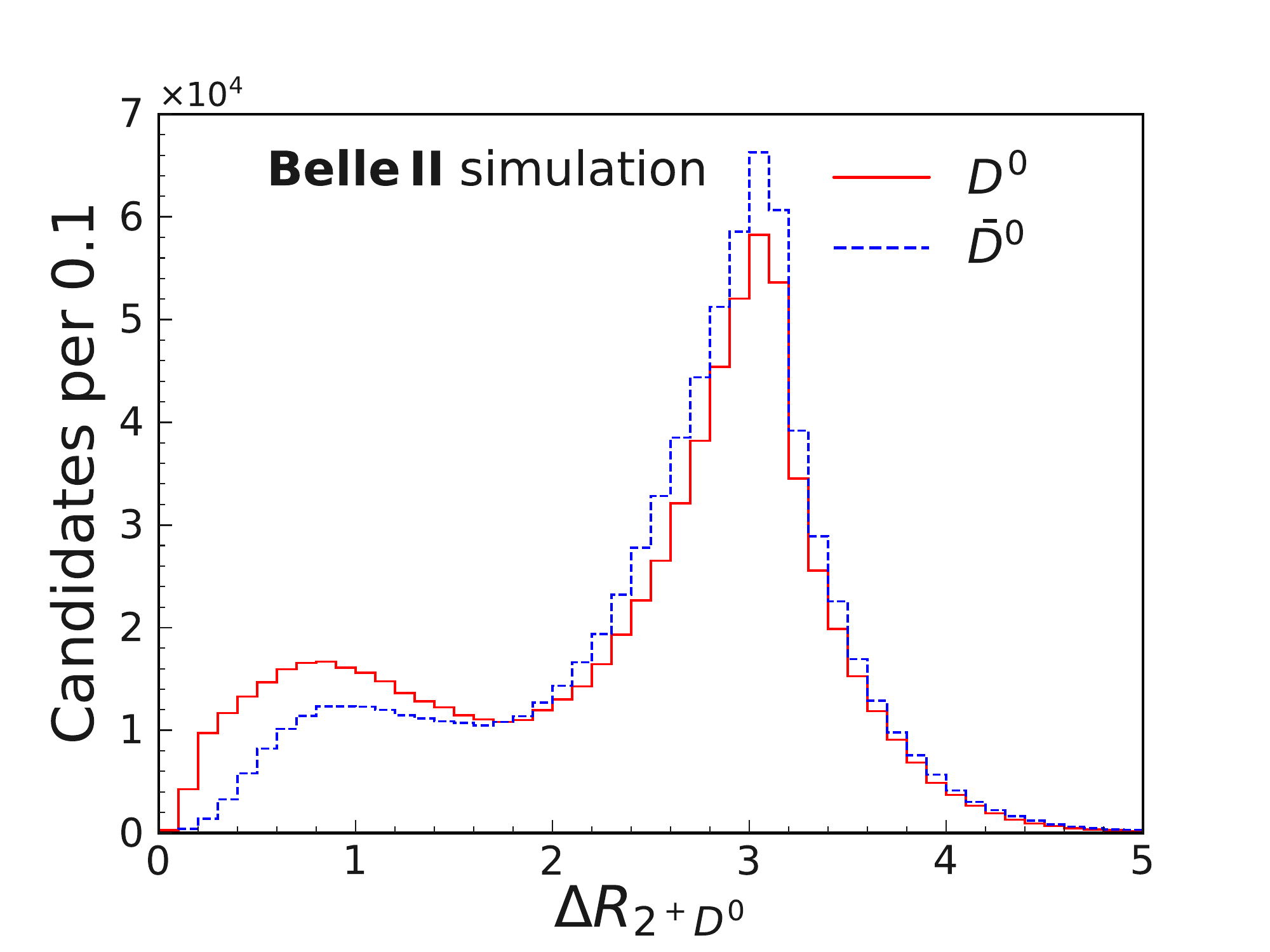}\hfil
\includegraphics[width=0.33\textwidth]{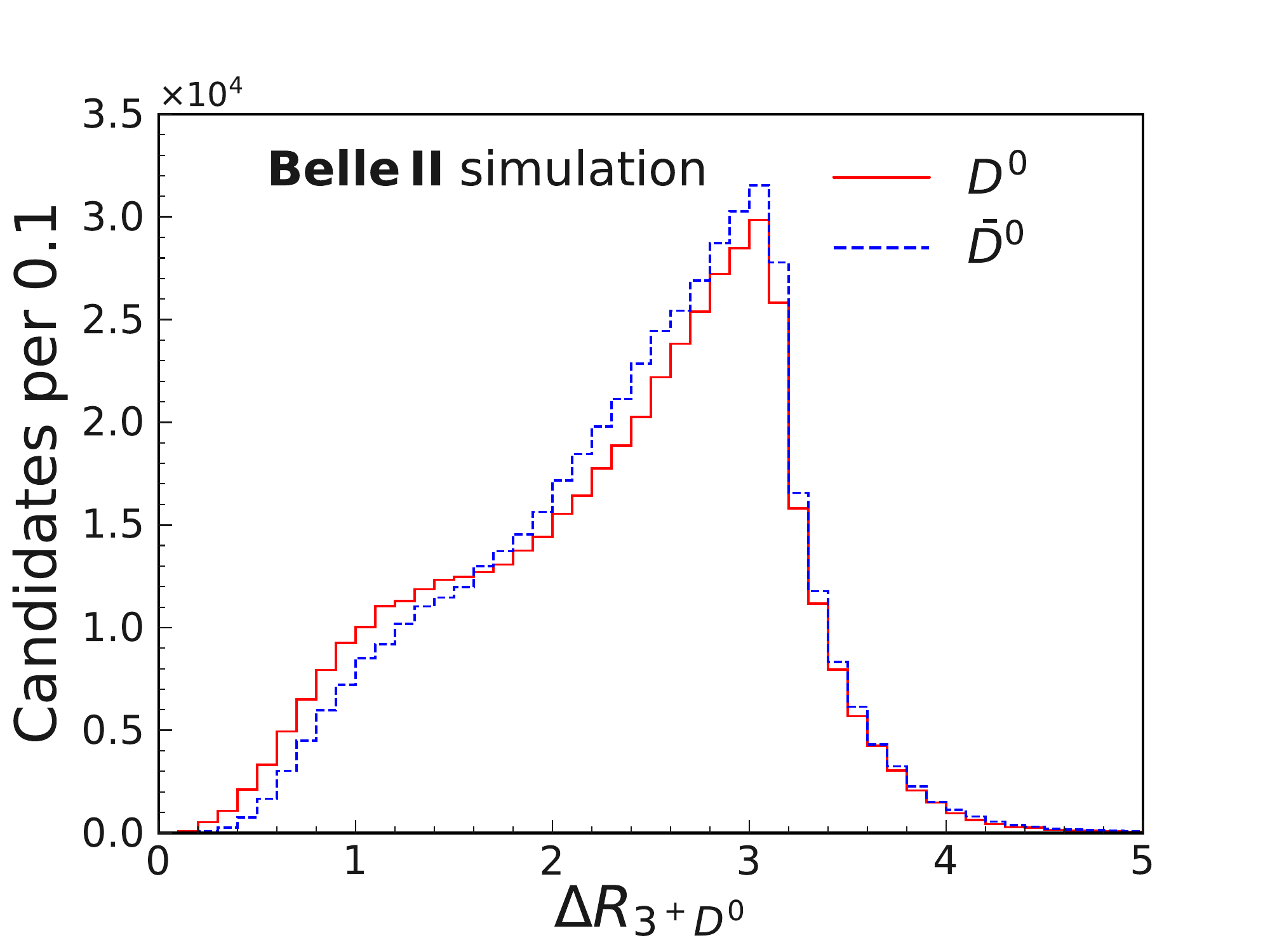}\\
\includegraphics[width=0.33\textwidth]{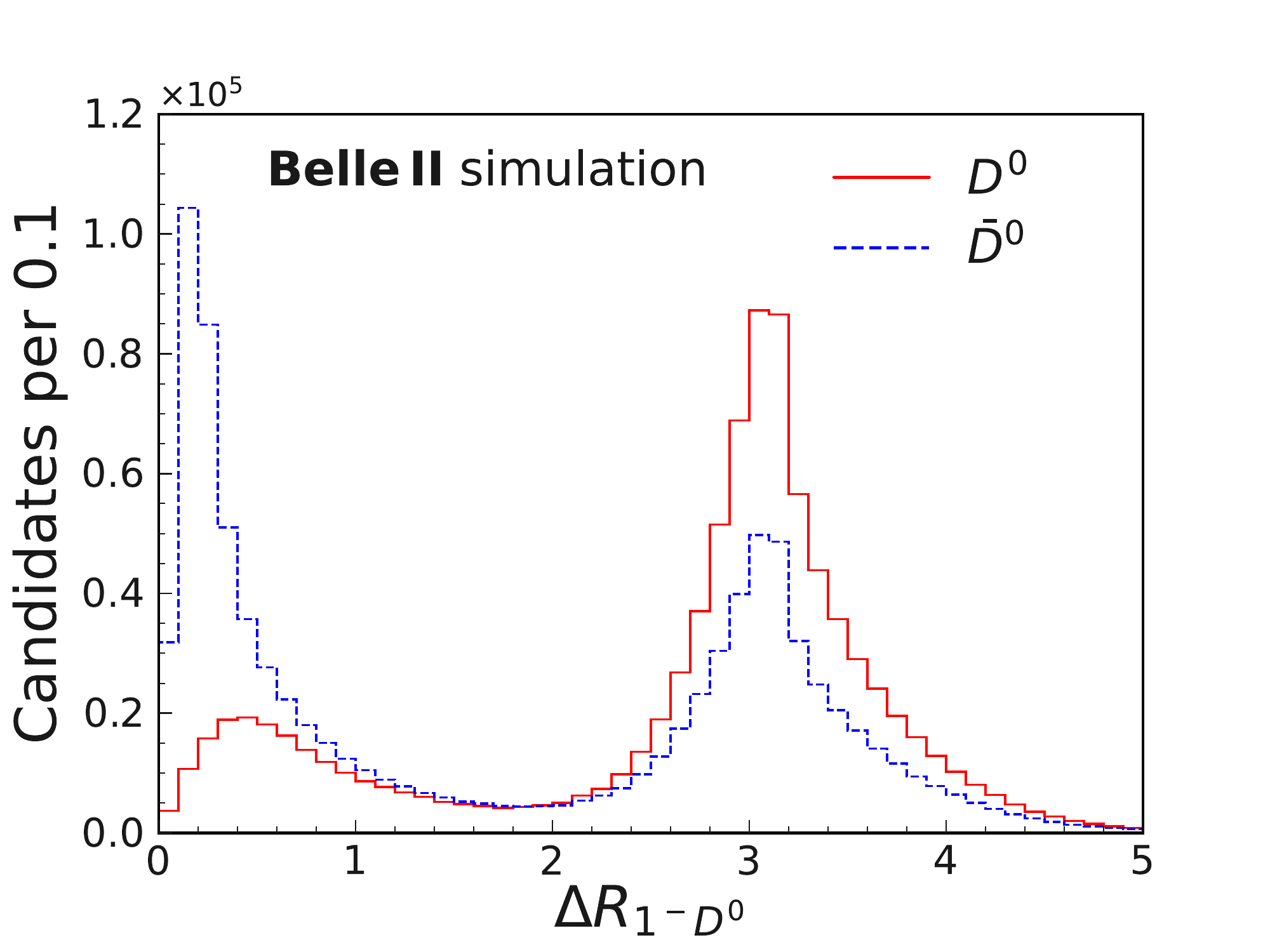}\hfil
\includegraphics[width=0.33\textwidth]{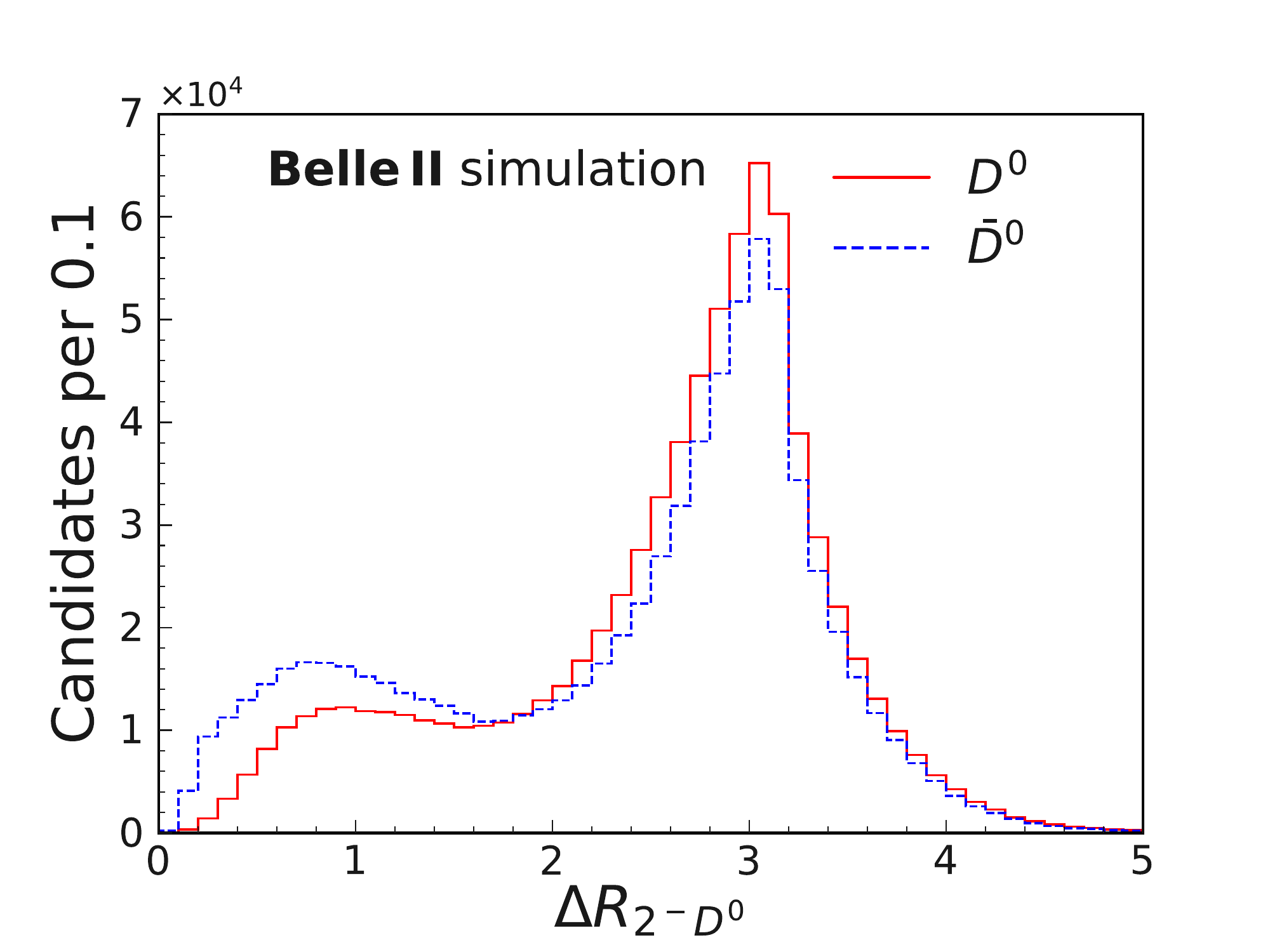}\hfil
\includegraphics[width=0.33\textwidth]{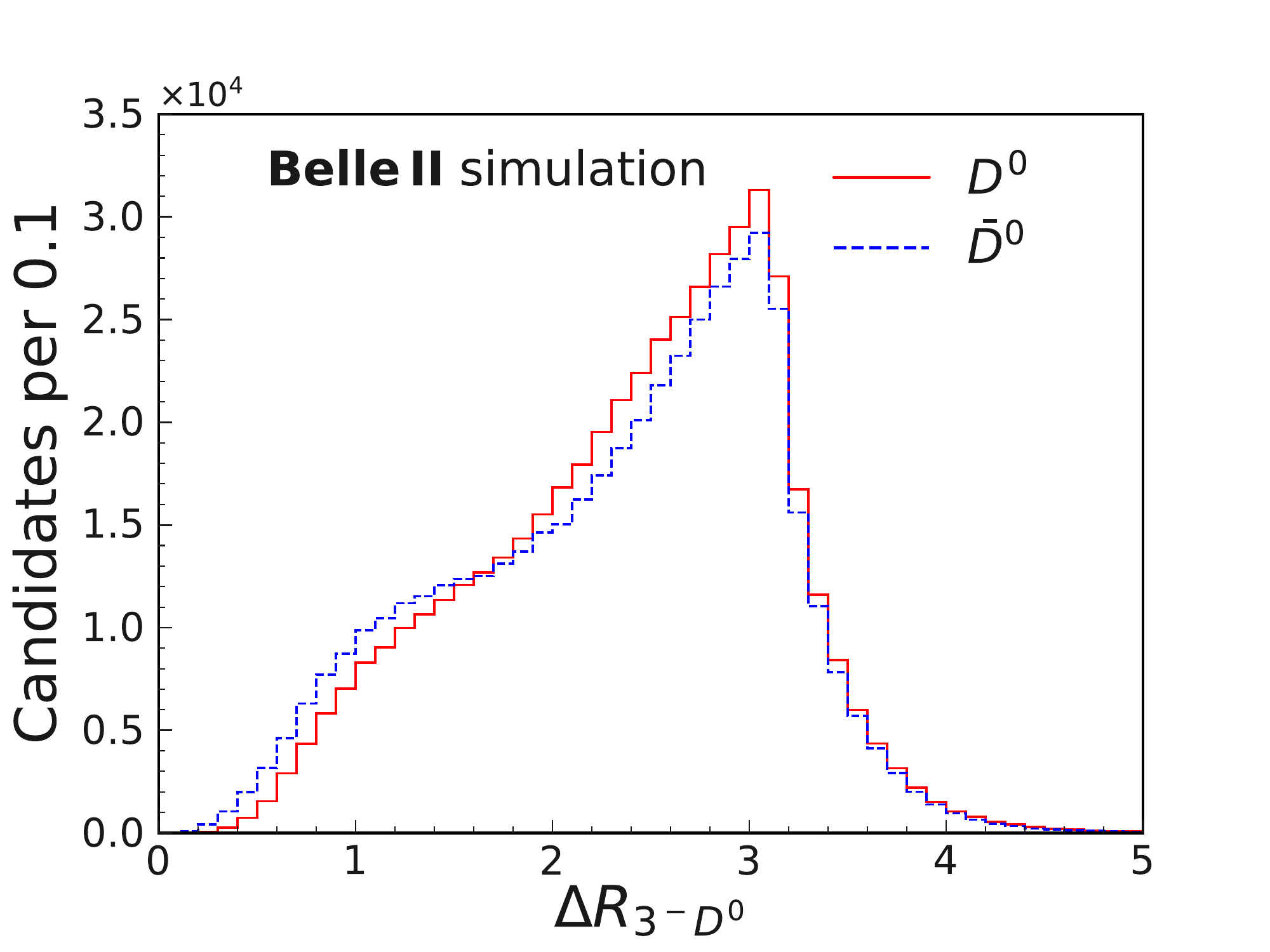}\\
\includegraphics[width=0.33\textwidth]{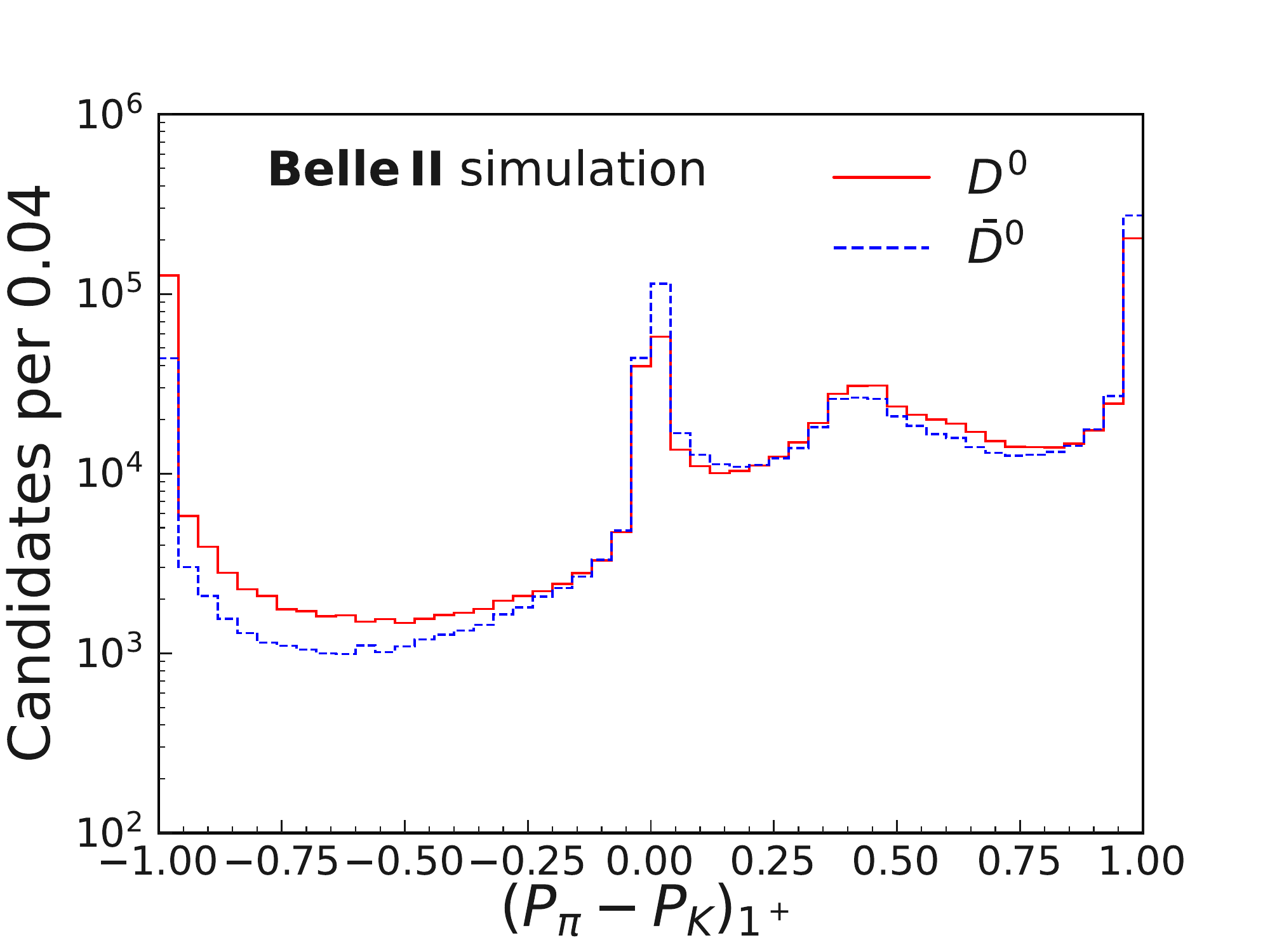}\hfil
\includegraphics[width=0.33\textwidth]{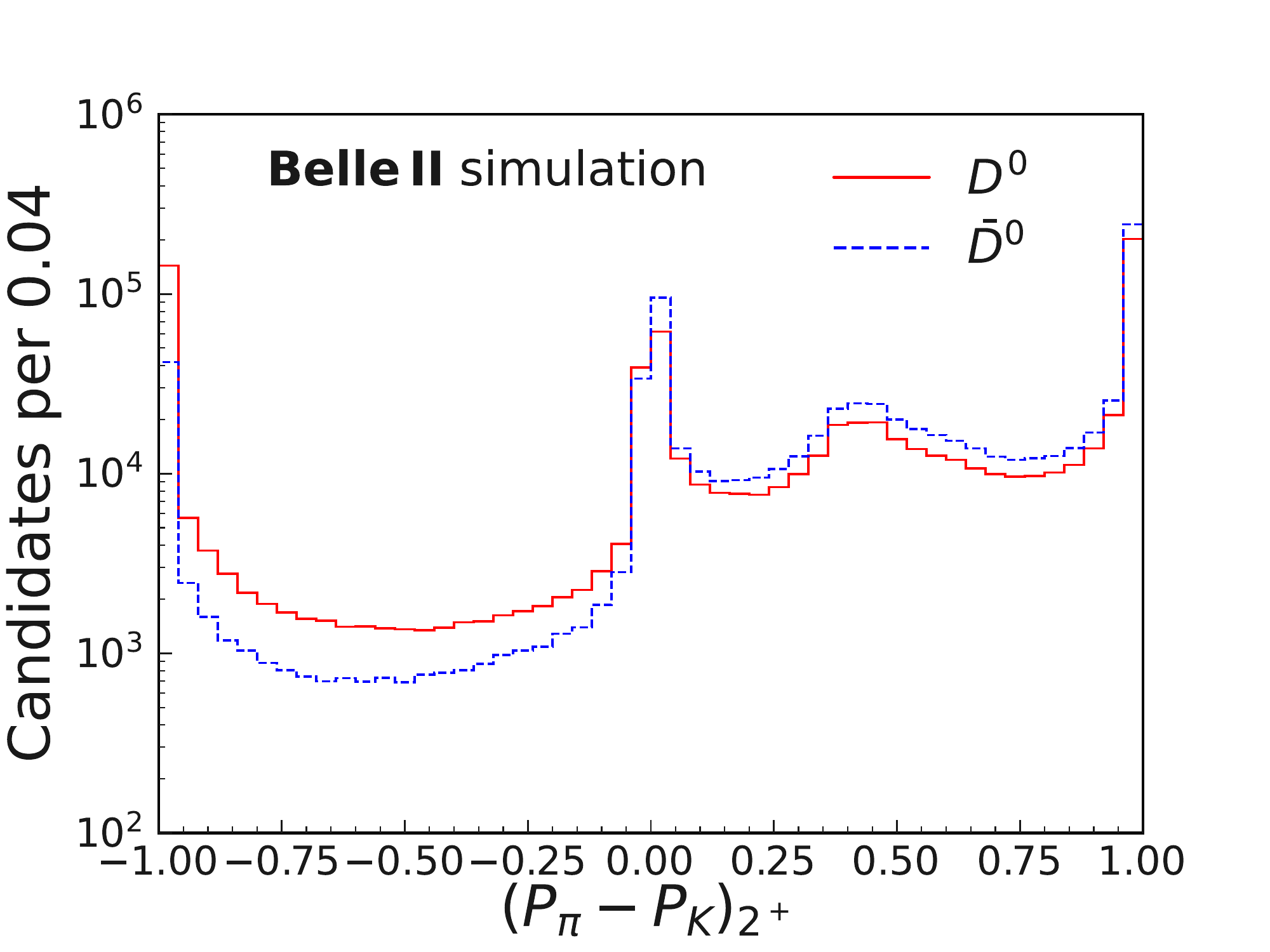}\hfil
\includegraphics[width=0.33\textwidth]{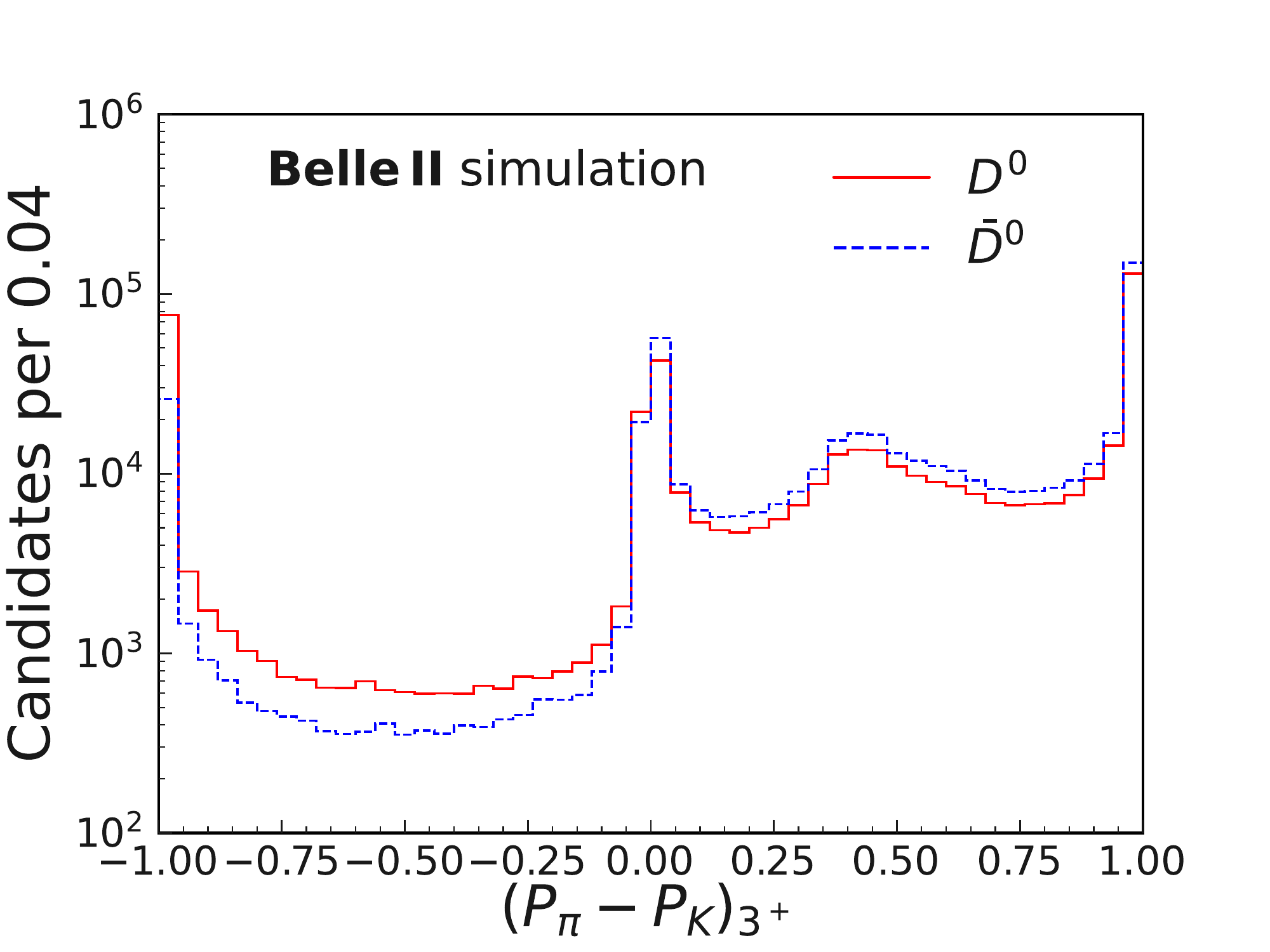}\\
\includegraphics[width=0.33\textwidth]{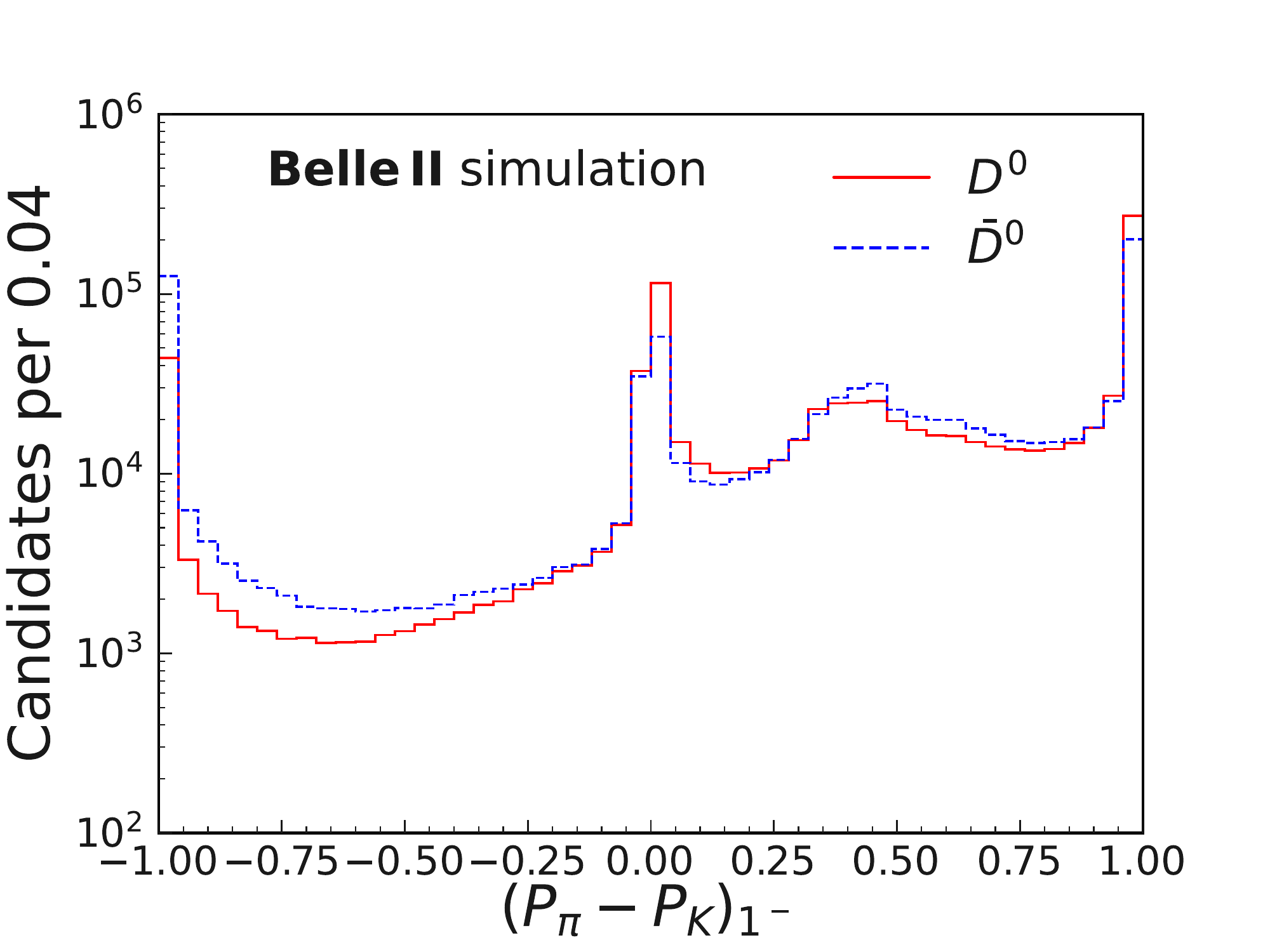}\hfil
\includegraphics[width=0.33\textwidth]{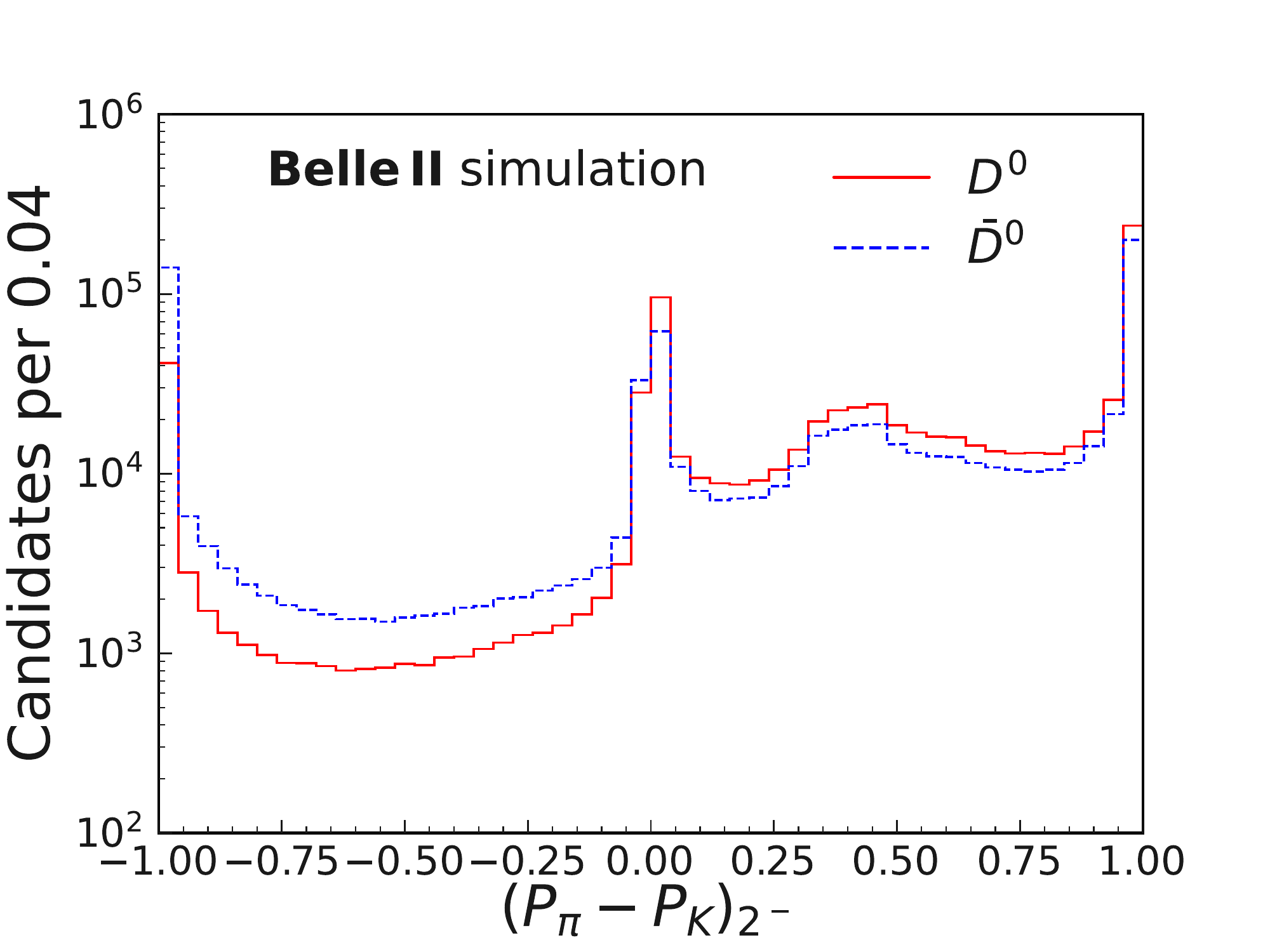}\hfil
\includegraphics[width=0.33\textwidth]{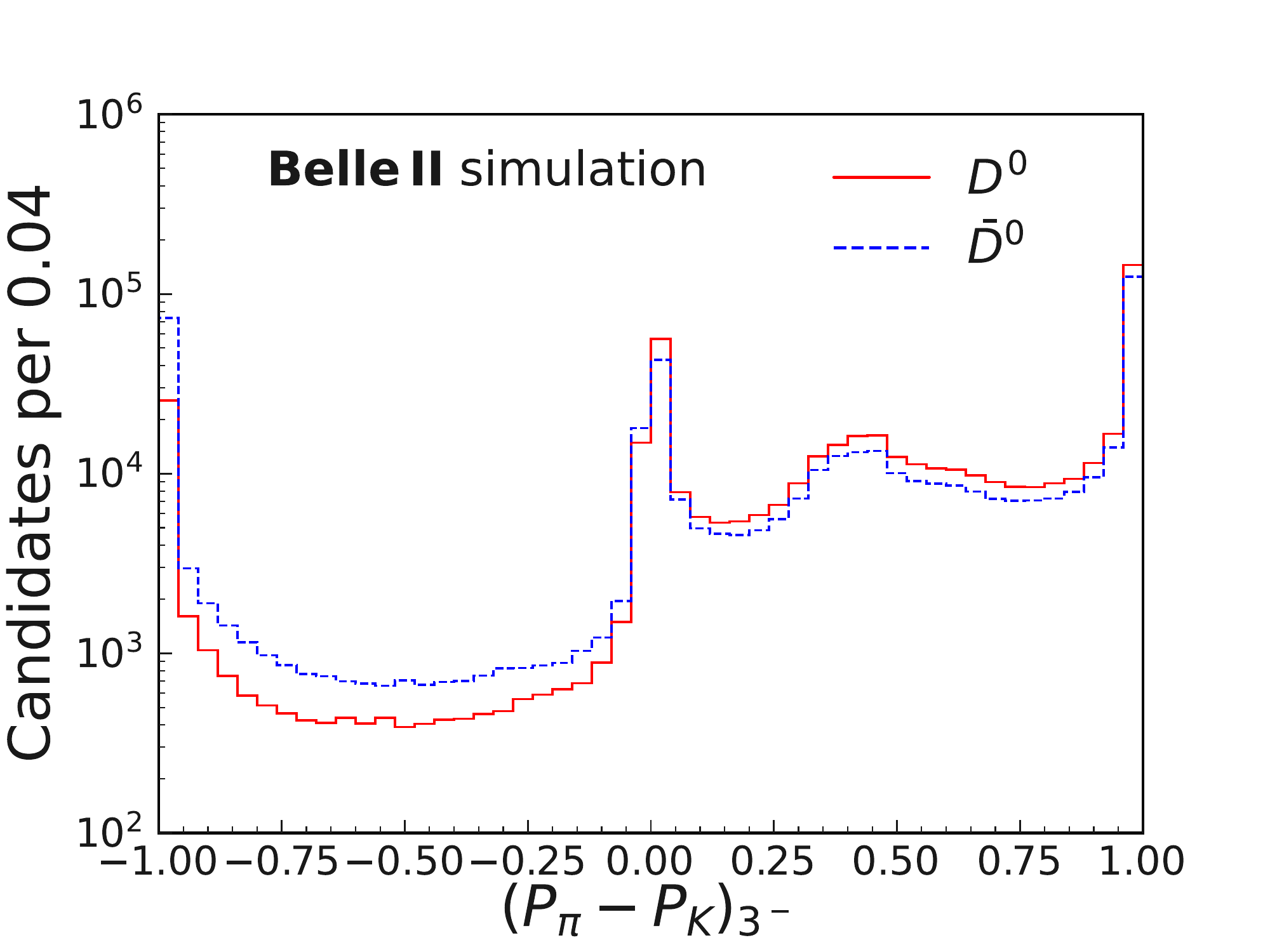}\\
\includegraphics[width=0.33\textwidth]{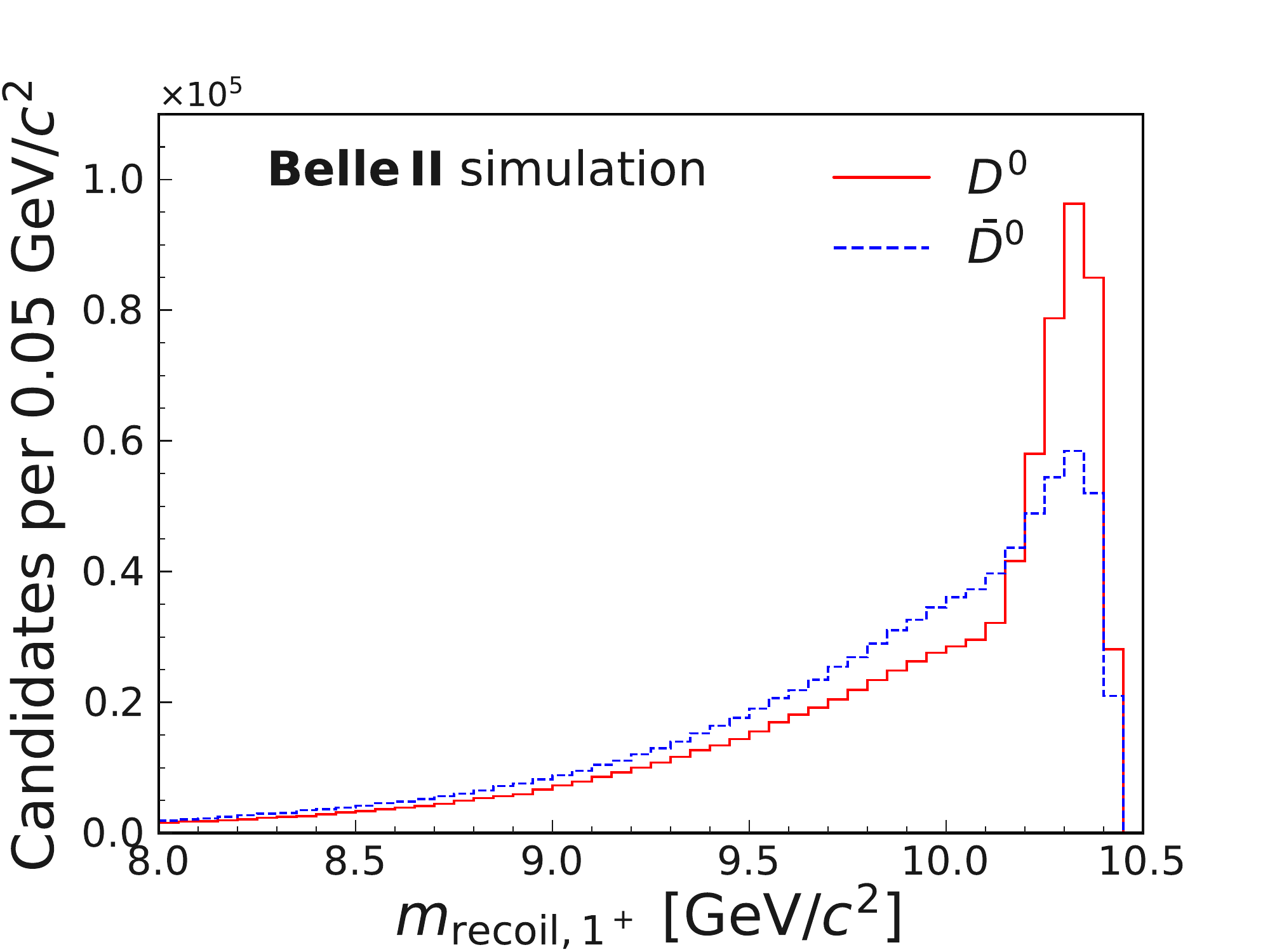}\hfil
\includegraphics[width=0.33\textwidth]{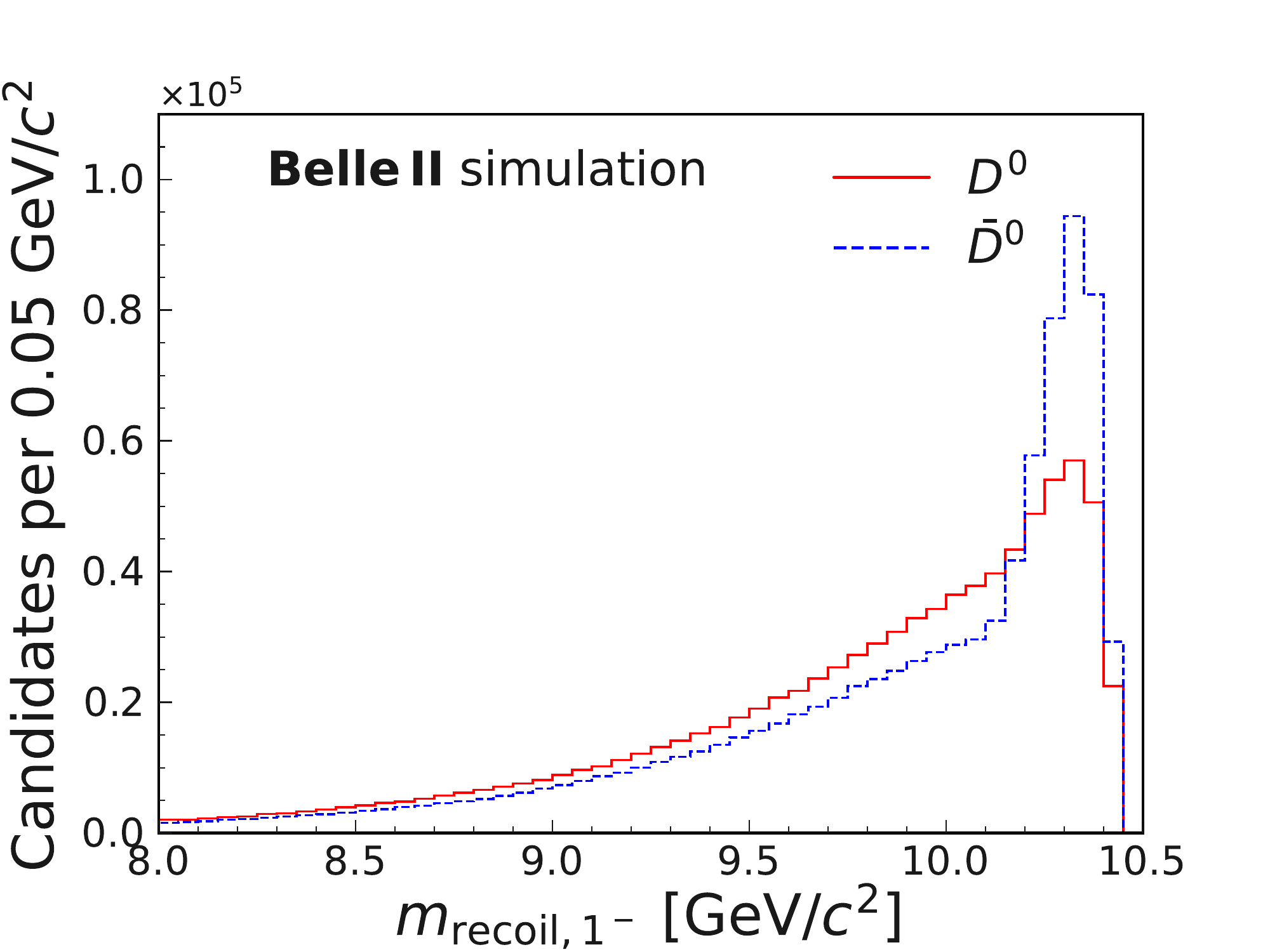}\\
\caption{Distributions of input variables separately for simulated \Dz and \Dzb true mesons. The angle $\Delta R$ is shown in the first and second row, the particle identification probabilities $P_\pi-P_K$ in the third and fourth row, and the recoil mass $m_{\rm recoil}$ in the last row. Electric charge and rank of the ROE particle are indicated in the subscript.\label{fig:tagger_input}}
\end{figure*}

\begin{figure}[t]
\centering
\includegraphics[width=0.45\textwidth]{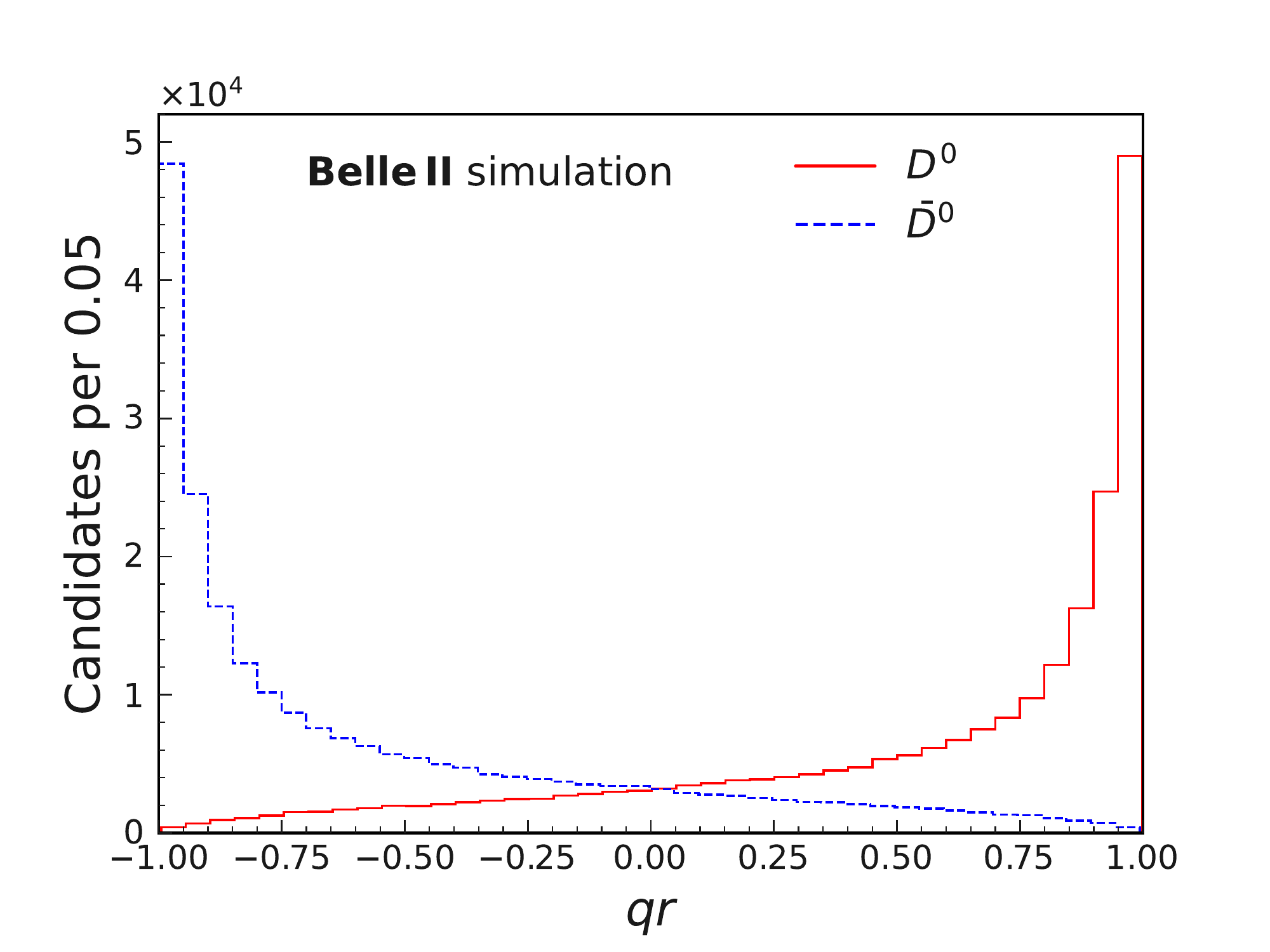}\\
\caption{Distributions of the predicted $qr$ for simulated \Dz and \Dzb true mesons in the testing sample.}\label{fig:tagger-output}
\end{figure}

The HBDT is trained using simulated $\Dz\to\neu\neub$ as signal, so that every reconstructed particle belongs to the ROE and cannot be associated to the signal meson, thus minimizing possible correlations of the CFT response with the signal decay mode. We train the algorithm with 1.35 million decays. \Cref{fig:tagger_input} shows the distributions of the input variables, separately for true \Dz and \Dzb mesons. The trained HBDT is tested on an independent sample of 450 thousand signal $\Dz\to\neu\neub$ decays to assess overtraining. During testing we consider only signal \Dz mesons that have at least one ROE particle fulfilling one of the categories of \cref{tab:tagging-categories} to have a consistent set of labels as used for training. The output of the HBDT on the test sample is shown in \cref{fig:tagger-output} separately for true \Dz and \Dzb mesons. Comparison of the signal flavor predicted by the CFT and the true label shows that the algorithm predicts the correct flavor in approximately $83\%$ of decays.

\begin{figure*}[ht]
\centering
\includegraphics[width=0.75\textwidth]{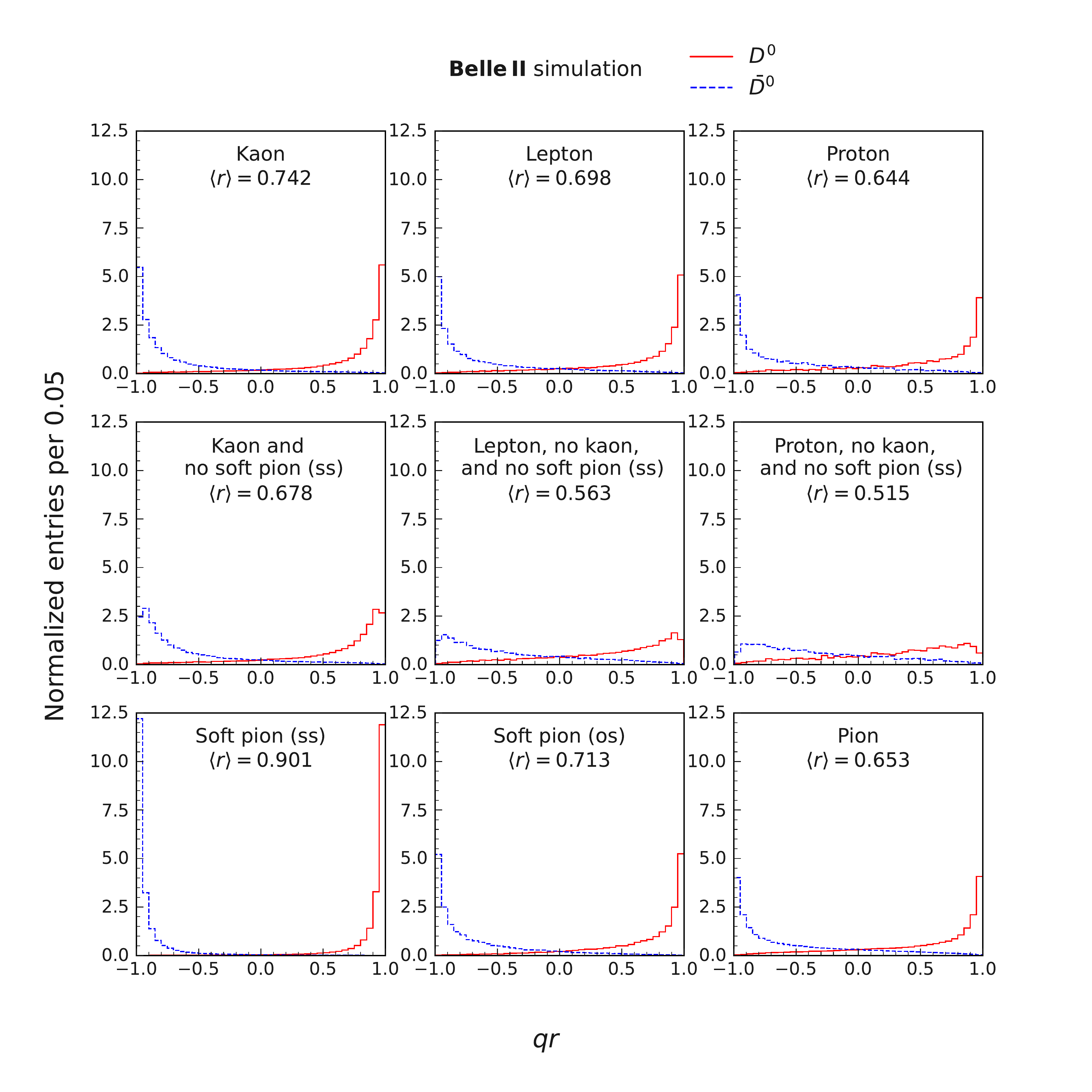}
\caption{Normalized distributions of the predicted $qr$ for simulated \Dz and \Dzb true mesons, and separately for subsets of the testing sample in which (top-left) one of the input ROE particle is a kaon tag, (top-center) one ROE particle is a lepton tag, (top-right) one ROE particle is a proton tag, (middle-left) one ROE particle is a kaon tag and there is no same-side (ss) soft-pion tag, (middle-center) one ROE particle is a lepton tag and there are no kaon or same-side soft-pion tags, (middle-right) one ROE particle is a proton tag and there are no kaon or same-side soft-pion tags, (bottom-left) one ROE particle is a same-side soft-pion tag, (bottom-center) one ROE particle is an opposite-side (os) soft-pion tag, and (bottom-right) one ROE particle is a pion tag. The average dilutions for the flavor-averaged and flavor-separated samples are reported.}\label{fig:tagger-split-by-categories}
\end{figure*}

\Cref{fig:tagger-split-by-categories} shows the CFT output for candidates selected using different criteria based on the known flavor tagging categories of \cref{tab:tagging-categories,tab:flavor-charge-relations}. The distributions clearly show that the CFT performs the best, with an average dilution $\mean{r}\approx0.90$, when a same-side soft-pion tag is present. The next best average dilution ($\mean{r}\approx0.75$) is obtained when a kaon tag is present. However, the performance decreases substantially ($\mean{r}\approx0.68$) when the presence of a kaon tag is not accompanied by the presence of a same-side soft-pion tag. The poorest performance ($\mean{r}\approx0.52$) is observed when a proton tag is present and there are no kaon or same-side soft-pion tags.

 %
\section{Performance\label{sec:performance}}
To evaluate the performance of the CFT and calibrate its prediction of the per-candidate dilution, we use data corresponding to an integrated luminosity of \lumi and collected by \belletwo from 2019 to 2022 at collision energies near the mass of the $\Upsilon(4S)$ resonance. The tagging performance is studied for neutral \D mesons and for \Dp and \Lc decays. Studying all three hadrons provides insight into the contributions from various tagging categories and enables validation of the results observed in simulation during testing and training.

We reconstruct the following signal decays of charmed hadrons: $\Dz\to\Km\pip$, $\Dz\to\Km\pip\pip\pim$, $\Dz\to\Km\pip\piz$, $\Dp\to\KS\pip$, $\Dp\to\Km\pip\pip$, and $\Lc\to p\Km\pip$. The neutral \D decays considered here proceed mainly through the Cabibbo-favored $c\to s$ transition. Hence, up to an $\order(10^{-3})$ contribution from \emph{wrong-sign} decays, the charge of the kaon identifies unambiguously the flavor of the neutral \D meson at the time of production. The kaon charge can be used as an approximation of the true tag and compared with the CFT decision to compute the mistag fraction in data. The \emph{wrong-sign} decays, which are a combination of doubly Cabibbo-suppressed $\Dzb\to\Km\pip(\pip\pim,\piz)$ decays and Cabibbo-favored $\Dz\to\Km\pip(\pip\pim,\piz)$ decays preceded by a $\Dzb\to\Dz$ oscillation, have an opposite correlation between the charge of the kaon and the production flavor of the \D meson. To account for such contributions we assume negligible \CP violation and correct the mistag fractions measured in data by subtracting the ratio of time-integrated wrong-sign to right-sign rates, which are $(3.798\pm0.014)\times10^{-3}$ for $\Dz\to K^\pm\pi^\mp$~\cite{LHCb:2017uzt}, $(3.33\pm0.27)\times10^{-3}$ for $\Dz\to K^\pm\pi^\mp\pip\pim$~\cite{LHCb:2016zmn}, and $(2.12\pm 0.13)\times10^{-3}$ for $\Dz\to K^\pm\pi^\pm\piz$ decays~\cite{Workman:2022ynf}.

\subsection{Reconstruction of signal decays\label{sec:reconstruction}}
The reconstruction of the signal candidates starts by selecting events that are inconsistent with Bhabha scattering and have at least three tracks with transverse momentum larger than $0.2\gevc$, transverse impact parameter smaller than $2\cm$ and longitudinal impact parameter smaller than $4\cm$.

Charged kaon, pion, and proton candidates are required to be in the acceptance of the drift chamber, originate from the \epem collision and be identified as kaons, pions, and protons, respectively, by requiring the particle identification probability for the given hypothesis to be larger than 0.9. Such requirements have typical efficiencies in the 60\%--80\% range and misidentification rates of a few percent, depending on the particle species. Candidate $\KS\to\pip\pim$ decays are formed by combining two oppositely charged particles under the pion mass hypothesis. A vertex fit is performed on the \KS candidates and the refitted mass is required to be in the range $[0.45,0.55]\gevcc$~\cite{Krohn:2019dlq}. Furthermore, the candidates' flight-length significance, defined as the ratio of flight length and its uncertainty $L_{\text{flight}}/\sigma_{\text{flight}}$, is computed from the results of the vertex fit and is required to be larger than 10.

We reconstruct photon candidates from localized energy deposits (clusters) in the electromagnetic calorimeter  that are consistent with an electromagnetic shower based on pulse-shape discrimination~\cite{Longo:2020zqt}. The cluster should have a polar angle within the acceptance of the drift chamber to ensure that it is not matched to tracks. It must include energy from at least two crystals and an energy deposit greater than 0.08\gev if located in the forward region ($12.4<\theta<31.4^\circ$), greater than 0.03\gev if in the barrel region ($32.2<\theta<128.7^\circ$), and greater than 0.06\gev if in the backward region ($130.7<\theta<155.7^\circ$). Two photon candidates are then combined to form a neutral pion candidate if the absolute difference in the azimuthal angles of the respective clusters is smaller than $86^\circ$, the corresponding opening angle is smaller than $80^\circ$, the invariant mass is in the range $0.120<M(\gamma\gamma)<0.145\gevcc$, and the momentum is larger than 0.4\gevc.

Combinations of charged kaon and pion candidates are used to form $\Dz\to\Km\pip$, $\Dz\to\Km\pip\pim\pip$,  and $\Dp\to\Km\pip\pip$ candidates. Charged kaon, pion, and neutral pion candidates are combined to form $\Dz\to\Km\pip\piz$ candidates. Candidate \KS mesons are combined with charged pions to form $\Dp\to\KS\pip$ candidates. Proton, kaon, and pion candidates are combined to form $\Lc\to p\Km\pip$ candidates. The \Dz, \Dp and \Lc candidates are required to have invariant masses in the ranges $[1.814,1.914]\gevcc$ ($[1.72,1.98]\gevcc$ for $\Dz\to\Km\pip\piz$),  $[1.819,1.919]\gevcc$ and $[2.248,2.323]\gevcc$, respectively. A vertex fit to the candidates is required to return $\chi^2$ probabilities in excess of 0.01. Charmed hadrons produced in \B-meson decays are rejected by requiring the center-of-mass momentum of the charmed hadron to be larger than 2.5\gevc.    

Given the large yield of these Cabibbo-favored decays, we expect the measurement of the tagging performance to be dominated by the systematic uncertainties. To facilitate further analysis, we therefore randomly discard 90\% of events when reconstructing a \Dz-meson decay, 20\% of the events when reconstructing a $\Dp\to\KS\pip$ decay, and 30\% of the events when reconstructing a $\Dp\to\Km\pip\pip$ decay. Events are not randomly discarded when reconstructing \Lc decays.

\begin{figure*}[ht]
    \centering
    \includegraphics[width=0.49\textwidth]{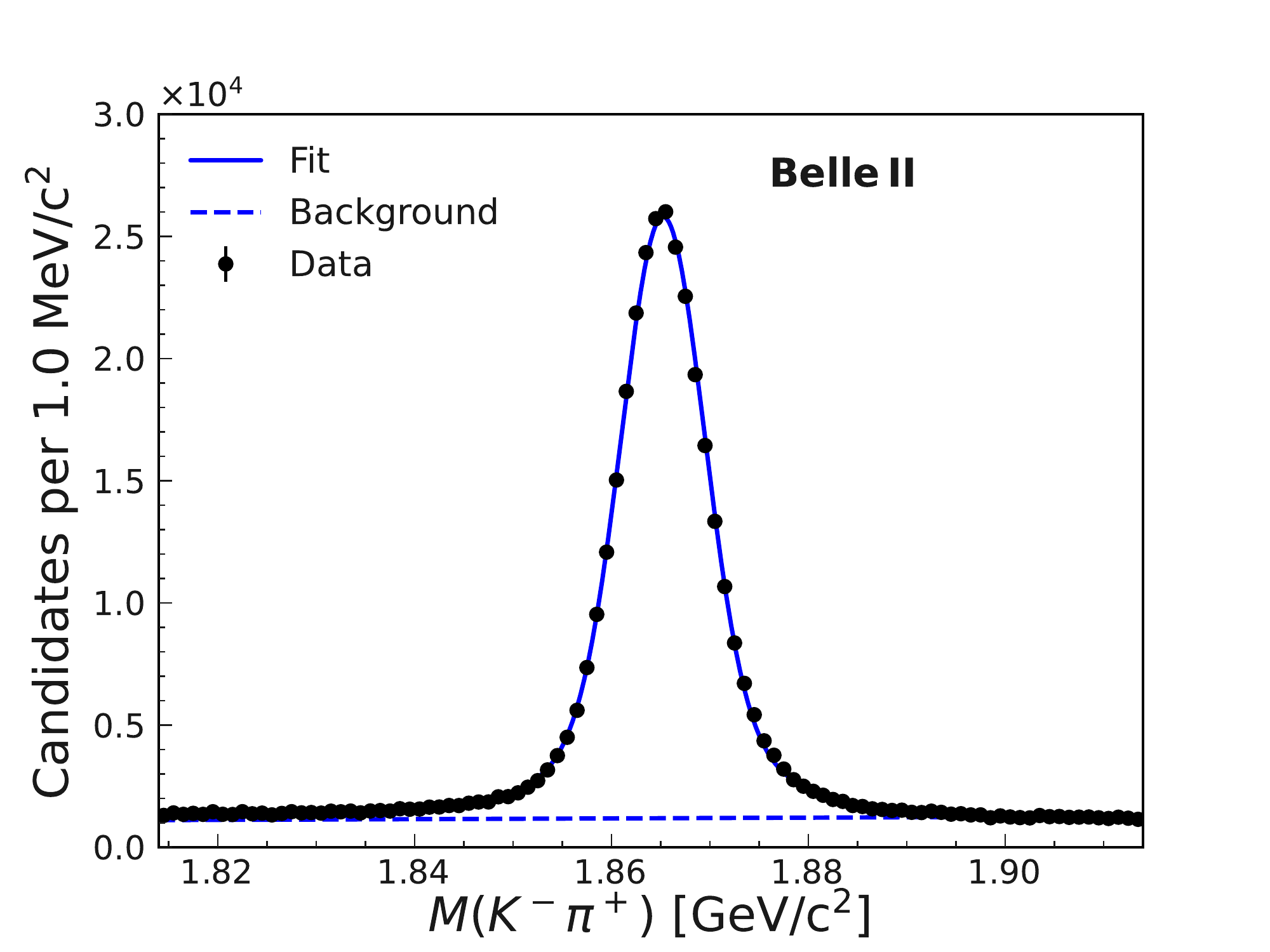}\hfil
    \includegraphics[width=0.49\textwidth]{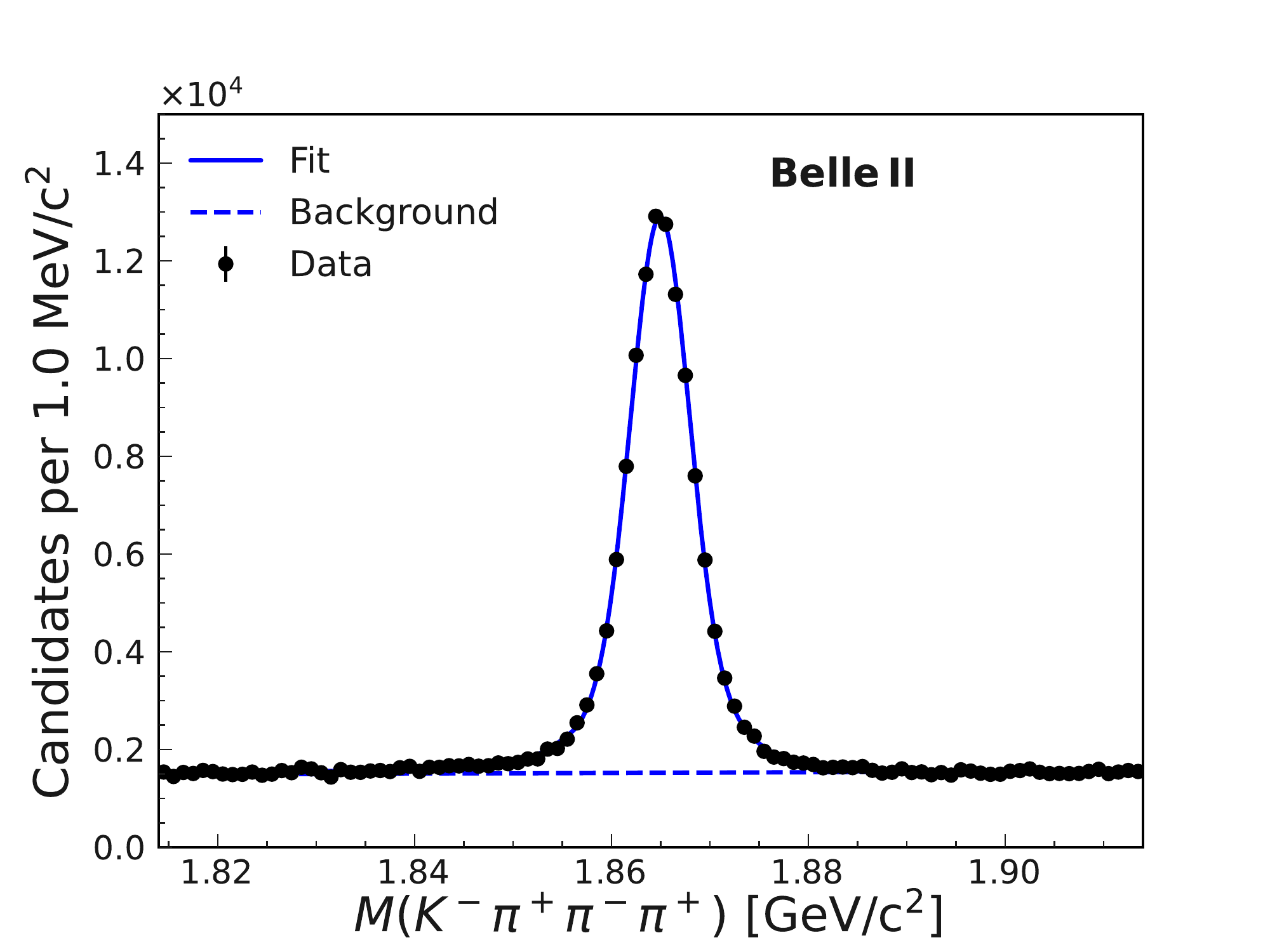}\\
    \includegraphics[width=0.49\textwidth]{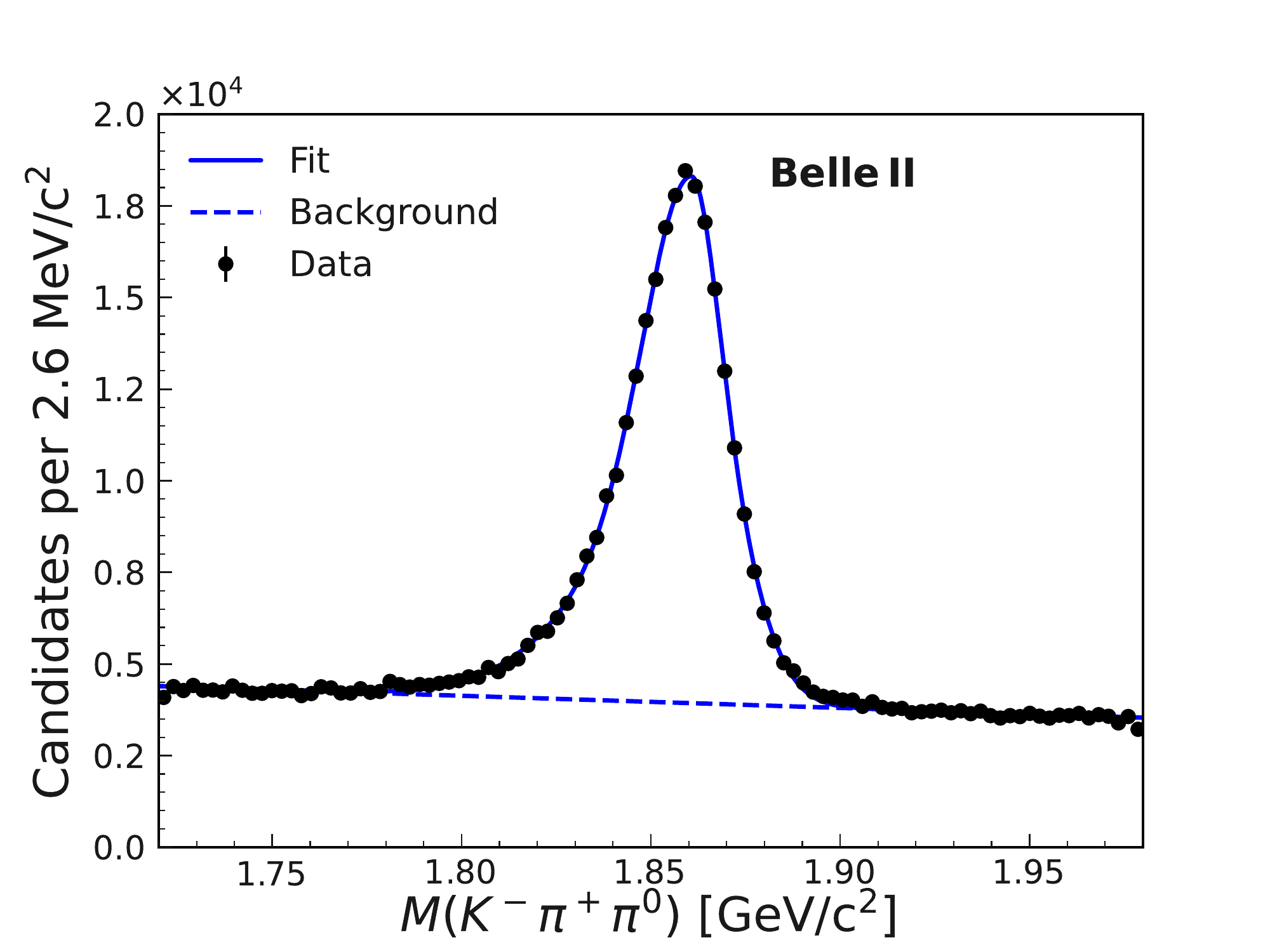}\hfil
    \includegraphics[width=0.49\textwidth]{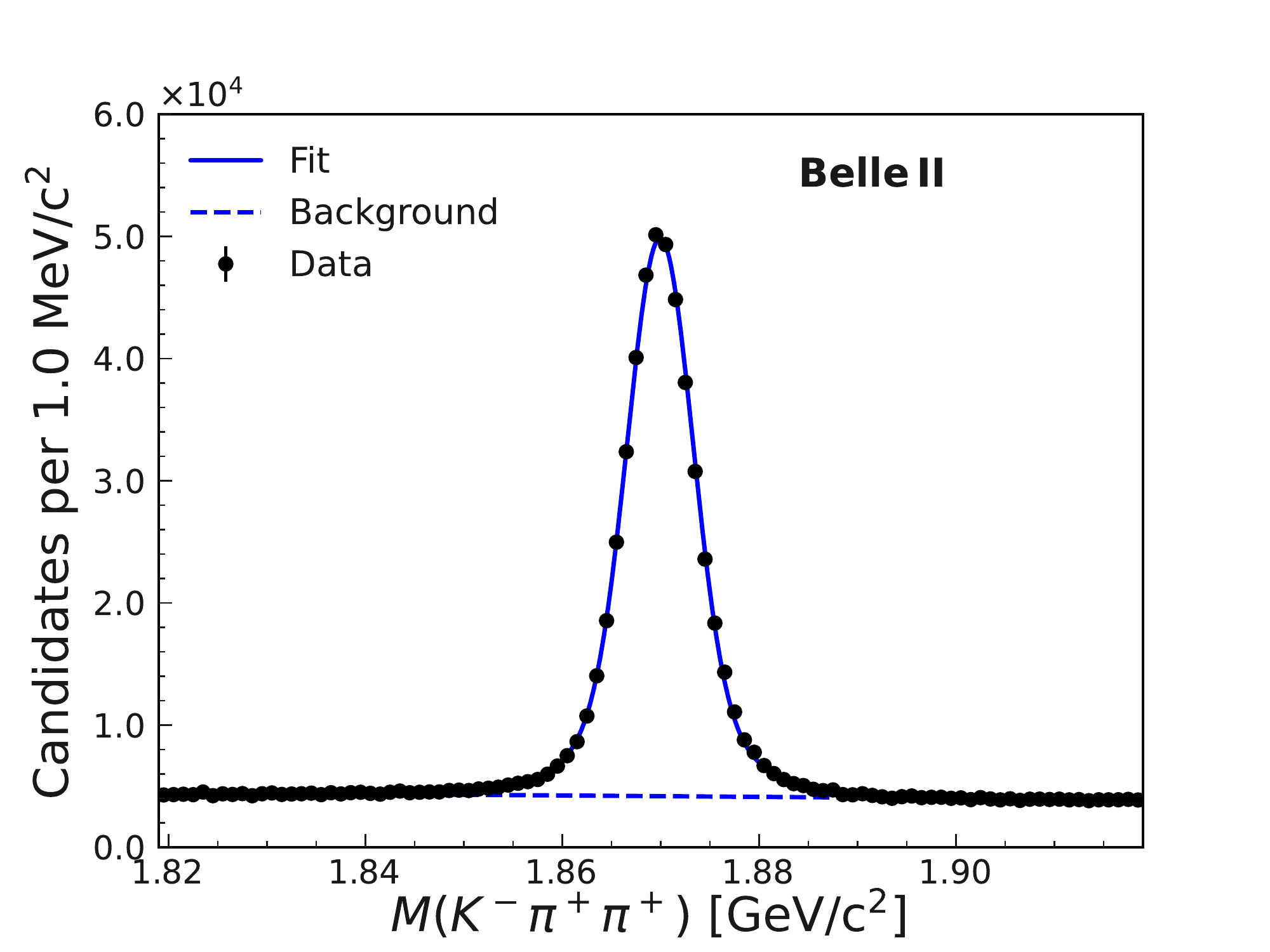}\\
    \includegraphics[width=0.49\textwidth]{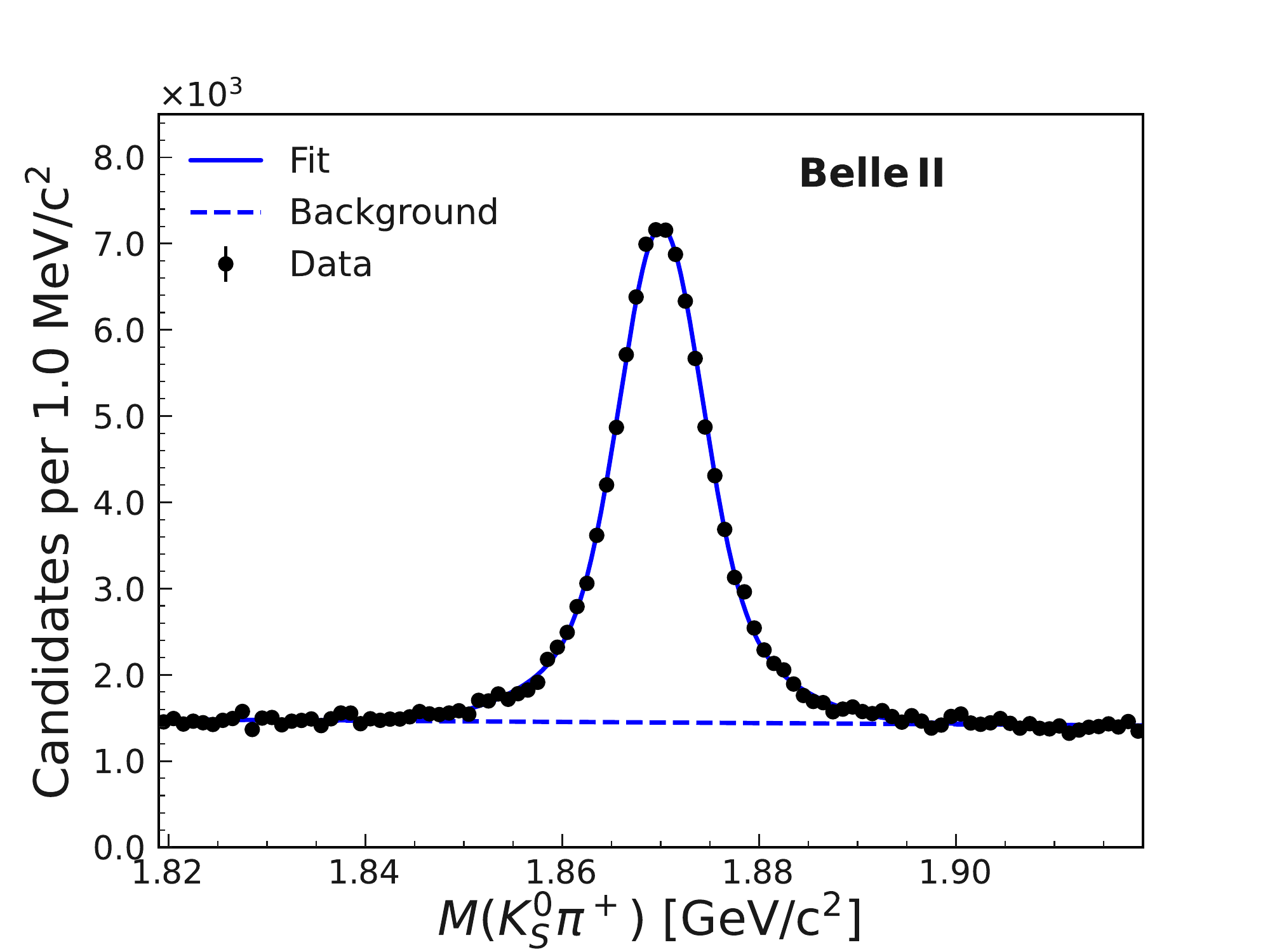}\hfil
    \includegraphics[width=0.49\textwidth]{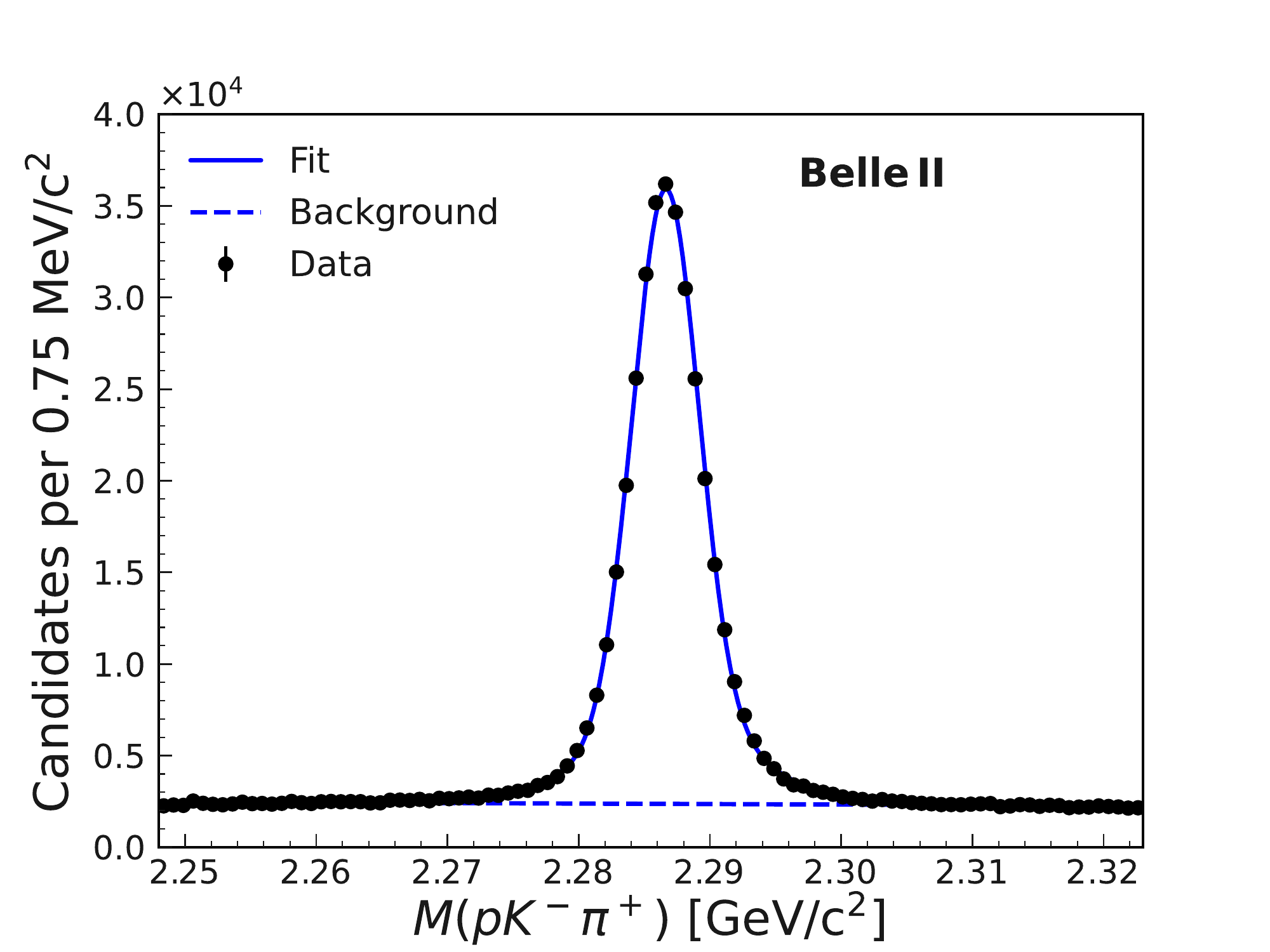}\\
    \caption{Mass distributions for selected (top left) $D^0\to K^-\pi^+$, (top right) $D^0\to K^-\pi^+\pi^-\pi^+$, (center left) $D^0\to K^-\pi^+\pi^0$, (center right) $\Dp\to K^-\pi^+\pi^+$, (bottom left) $\Dp\to\KS\pip$, and (bottom right) $\Lc\to p\Km\pip$ candidates in data with fit projections overlaid.}
    \label{fig:perf_fit}
\end{figure*}

To determine the tagging performance on signal-only decays, we fit to the mass distributions of the selected candidates (\cref{fig:perf_fit}) and use the \sPlot method~\cite{Pivk:2004ty} to subtract the background. In the fits, each signal peak is described by the sum of a Gaussian distribution and a Crystal Ball function~\cite{Gaiser:Phd,Skwarnicki:1986xj}; the background components by second-order polynomials. The fits estimate $300500\pm750$ $\Dz\to\Km\pip$, $102300\pm1400$ $\Dz\to\Km\pip\pip\pim$, $190500\pm730$ $\Dz\to\Km\pip\piz$, $75420\pm910$ $\Dp\to\KS\pip$, $438700\pm930$ $\Dp\to\Km\pip\pip$, and $330500\pm810$ $\Lc\to p\Km\pip$ signal decays.

\subsection{Efficiency and mistag rate\label{sec:perf_res}}
When estimating the performance we account for the possibility that the tagging efficiency, \efftag, and the mistag rate, \mistag, can be different for charm and anticharm flavors due to charge-asymmetries in detection and reconstruction. We therefore define the differences
\begin{equation}
\defftag=\efftag(\qtrue=+1) - \efftag(\qtrue=-1)
\end{equation}
and
\begin{equation}
\dmistag=\mistag(\qtrue=+1) - \mistag(\qtrue=-1),
\end{equation}
where \qtrue indicates the true flavor. As an example, in a sample of signal neutral \D mesons $\mistag(\qtrue=+1)$ is the fraction of \Dz mesons that are incorrectly classified as \Dzb and $\mistag(\qtrue=-1)$ is the fraction of \Dzb mesons that are incorrectly classified as \Dz. The tagging efficiency and mistag rates of \cref{eq:efftag,eq:mistag} correspond to the arithmetic averages $[\efftag(\qtrue=+1)+\efftag(\qtrue=-1)]/2$ and $[\mistag(\qtrue=+1)+\mistag(\qtrue=-1)]/2$, respectively.

\begin{figure*}[ht]
    \centering
    \includegraphics[width=0.49\textwidth]{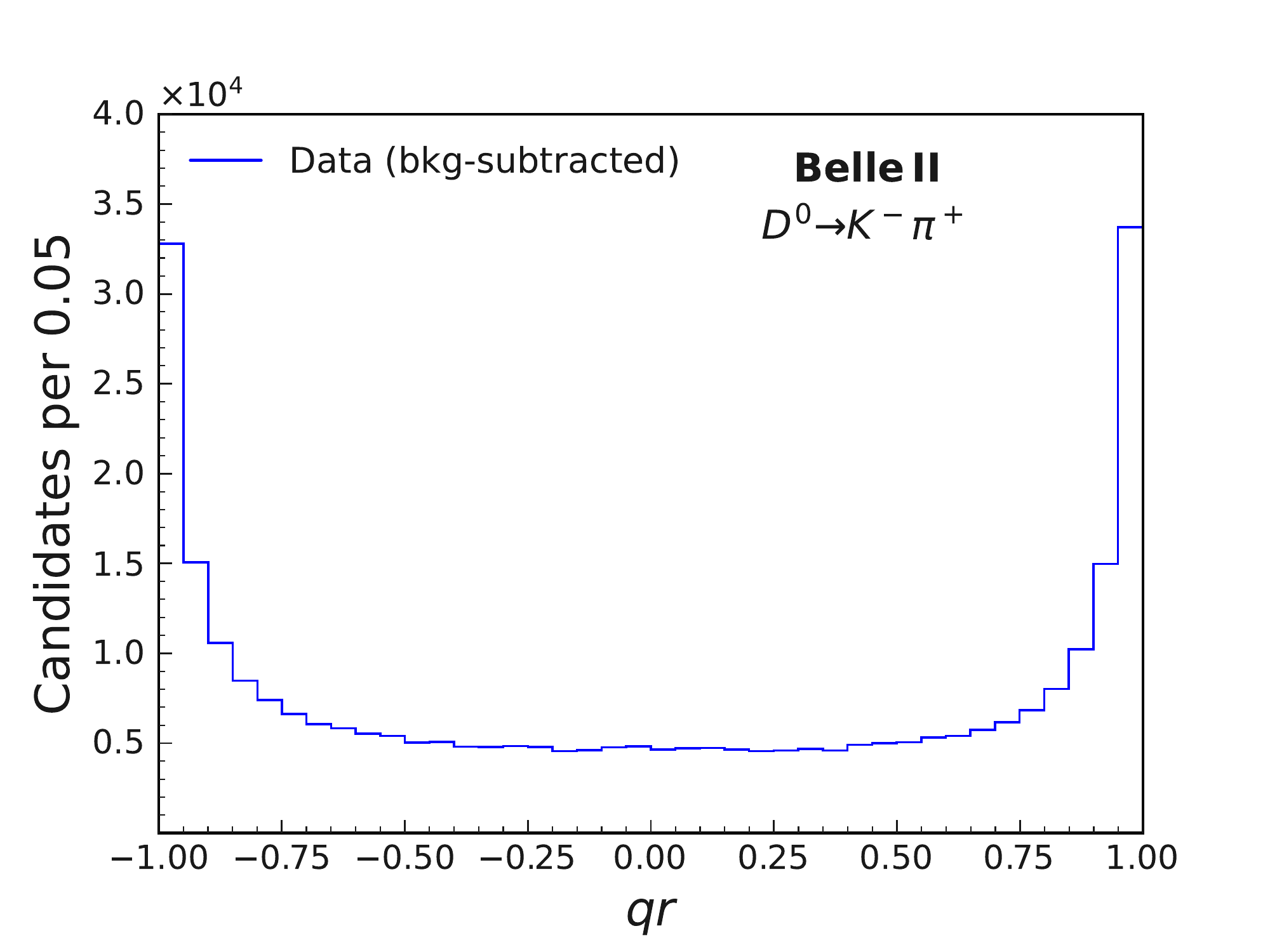}\hfil
    \includegraphics[width=0.49\textwidth]{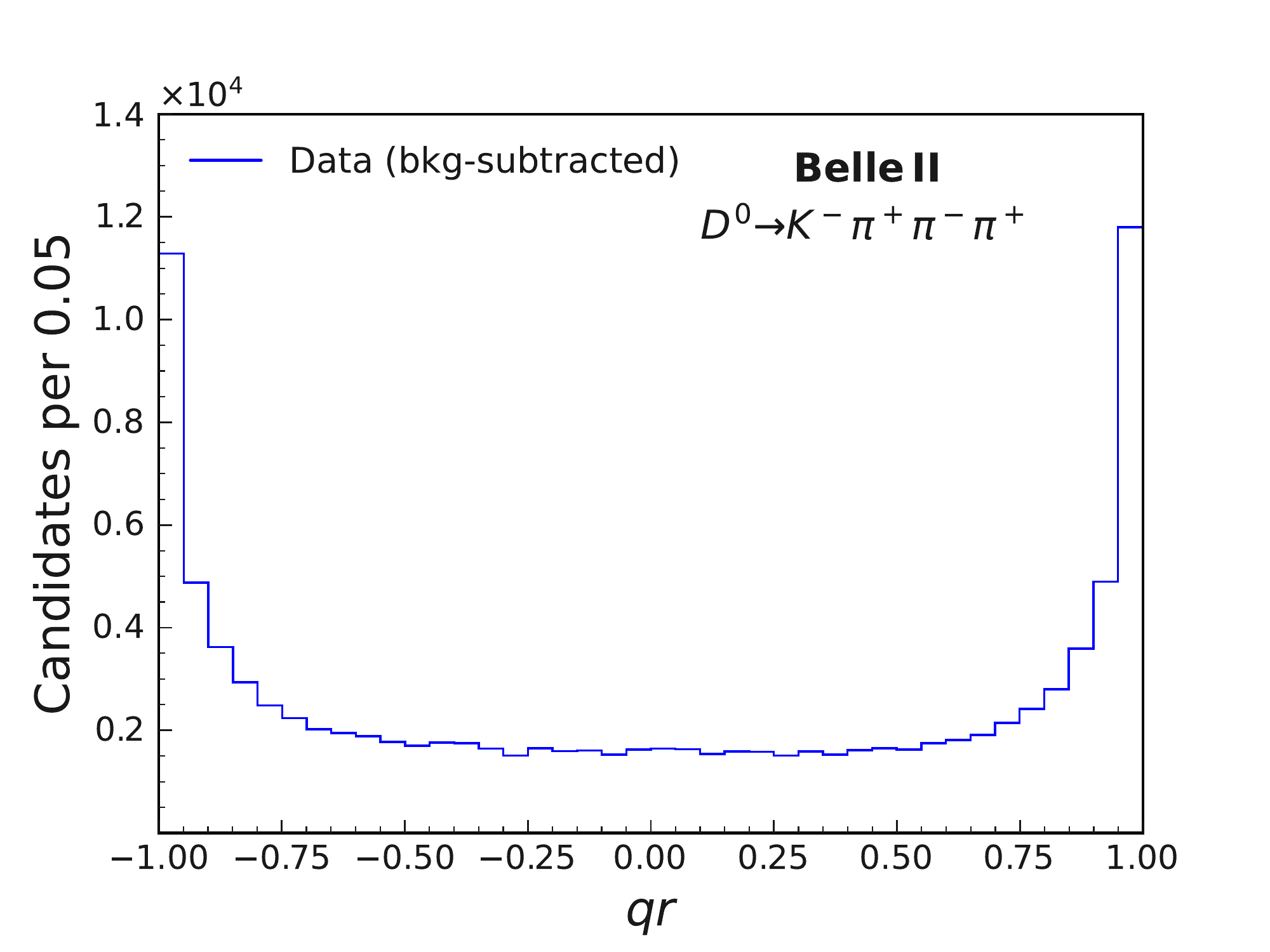}\\
    \includegraphics[width=0.49\textwidth]{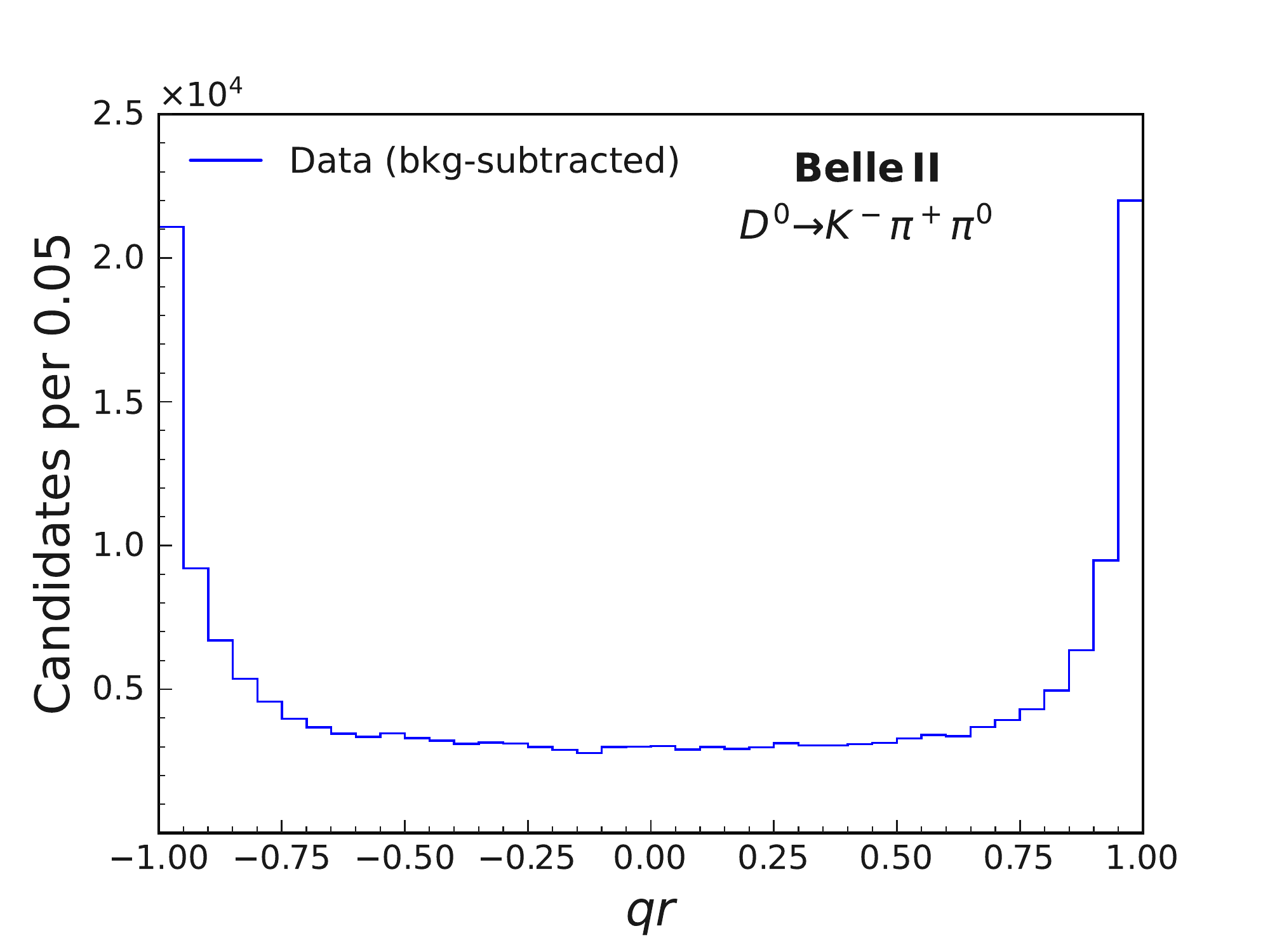}\hfil
    \includegraphics[width=0.49\textwidth]{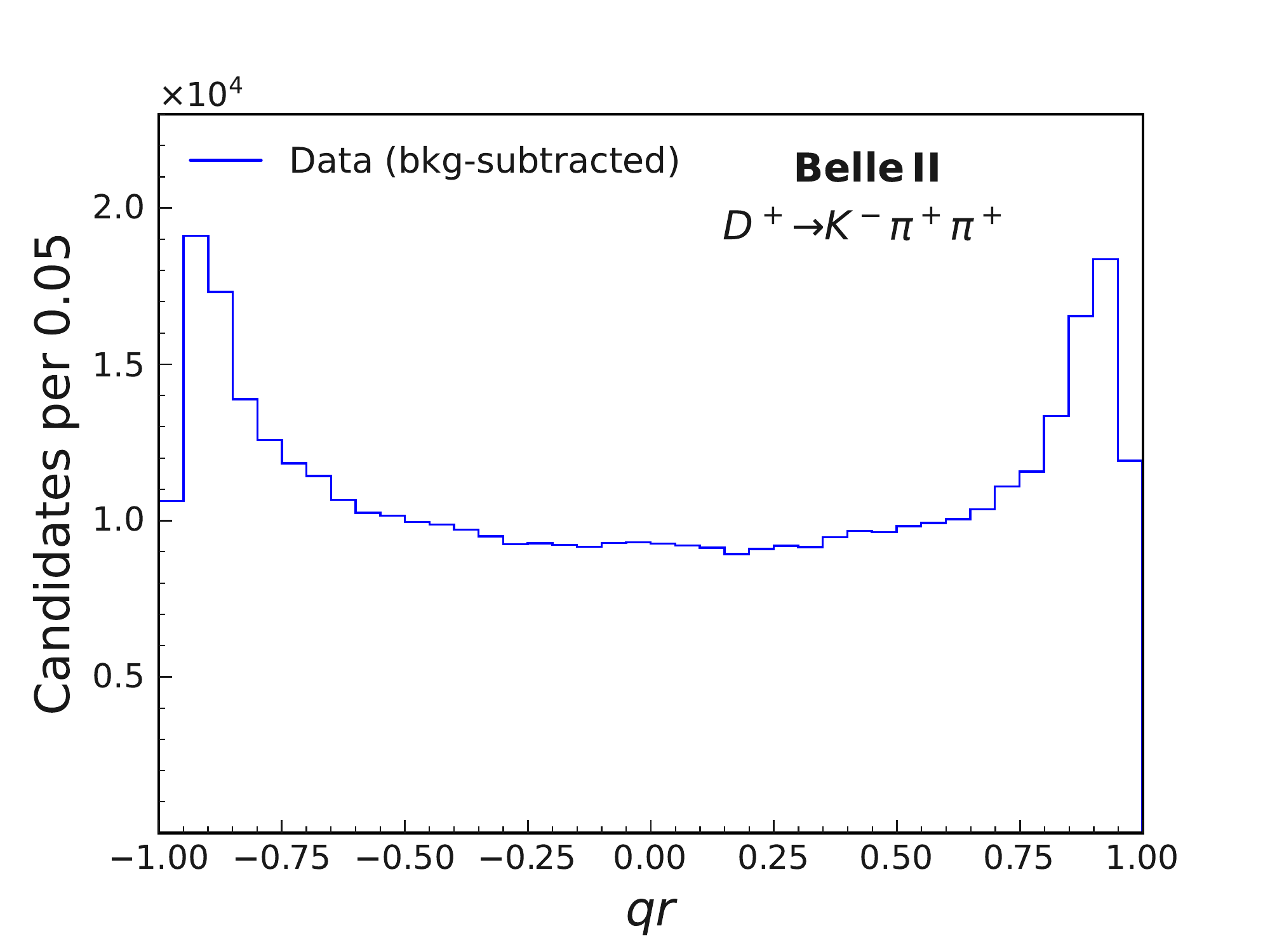}\\
    \includegraphics[width=0.49\textwidth]{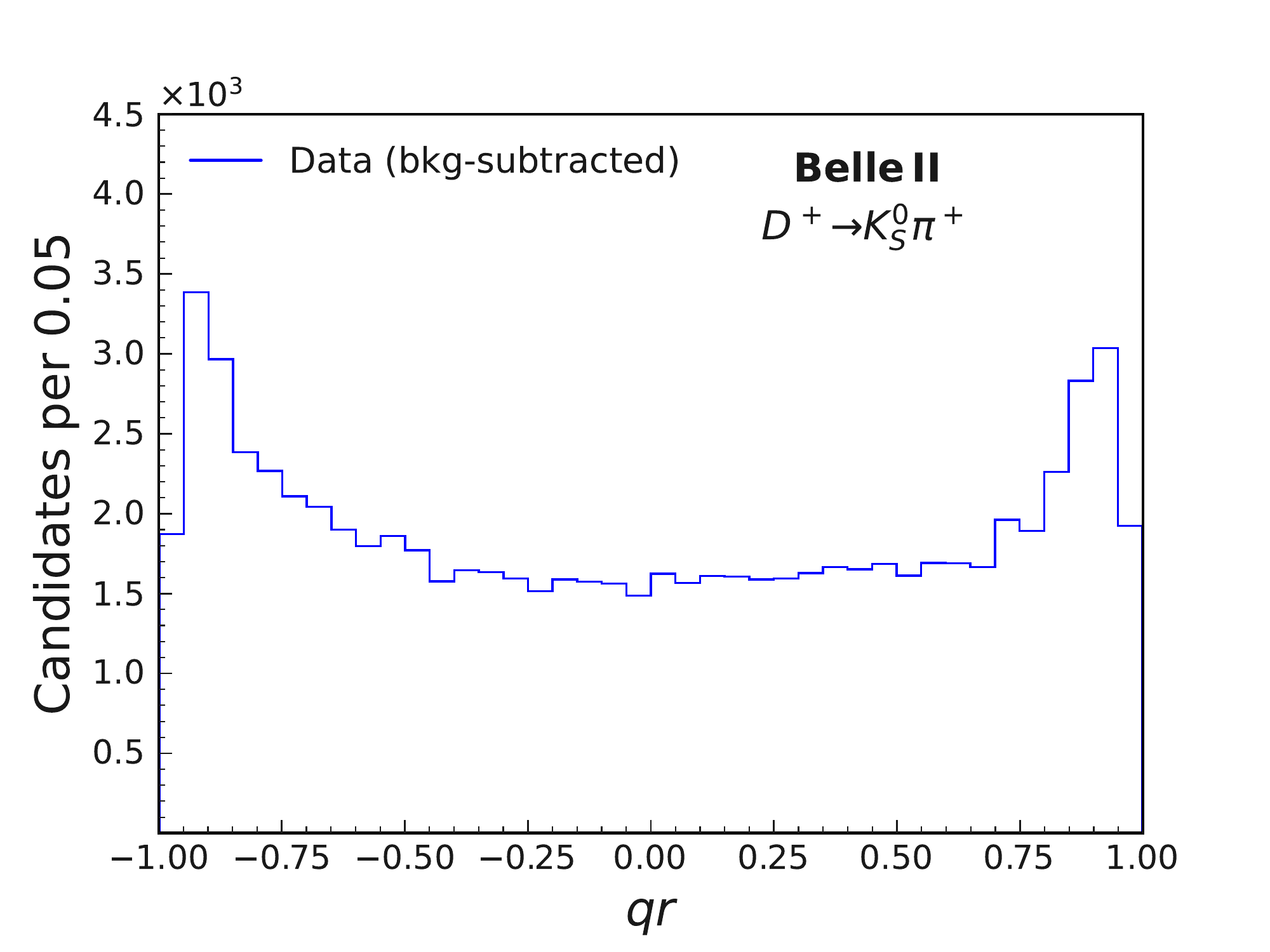}\hfil
    \includegraphics[width=0.49\textwidth]{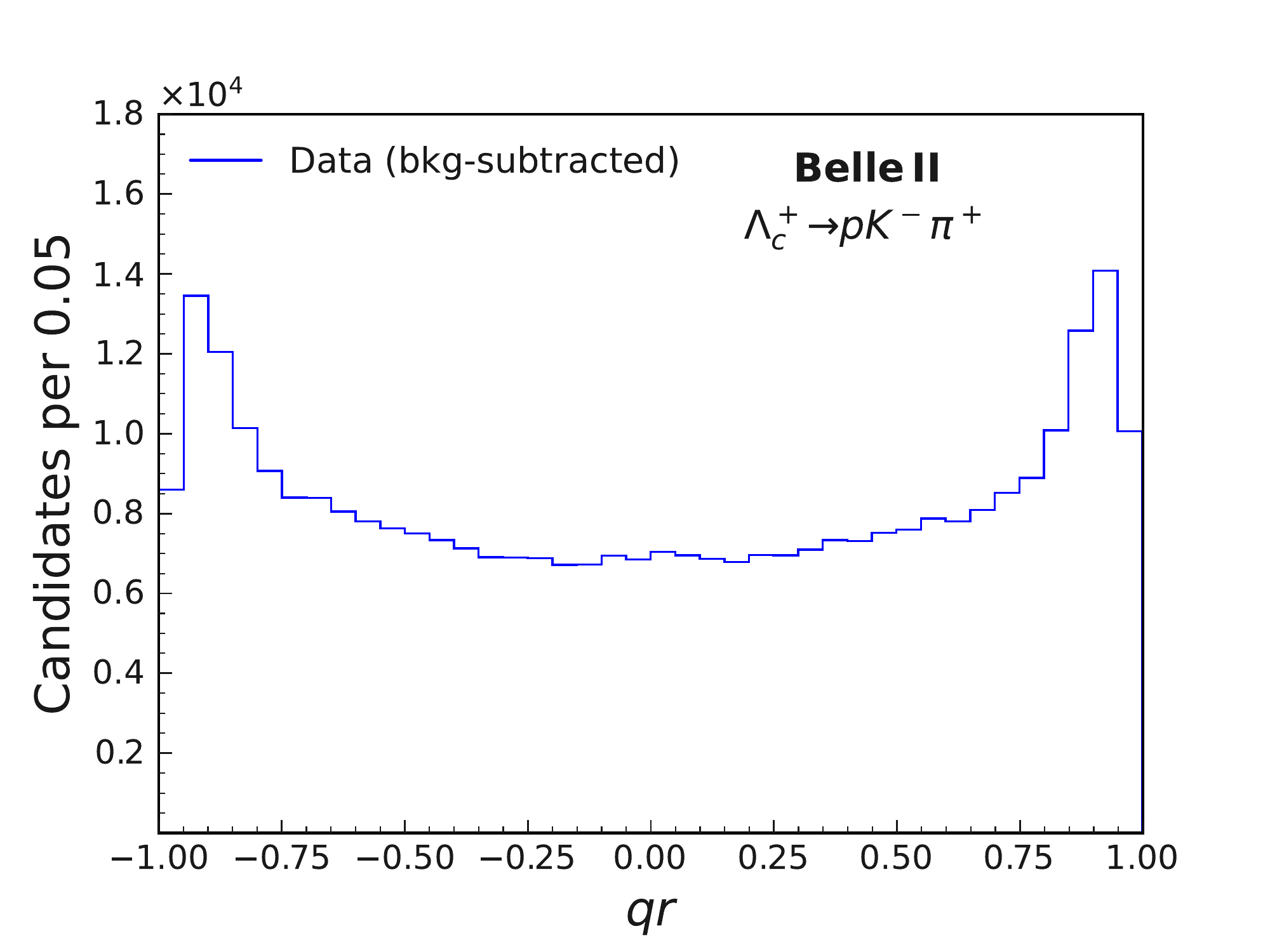}\\
    \caption{Distributions of the predicted $qr$ for background-subtracted (top left) $D^0\to K^-\pi^+$, (top right) $D^0\to K^-\pi^+\pi^-\pi^+$, (center left) $D^0\to K^-\pi^+\pi^0$,  (center right) $\Dp\to K^-\pi^+\pi^+$, (bottom left) $\Dp\to\KS\pip$, and (bottom right) $\Lc\to p\Km\pip$ decays in data.}
    \label{fig:perf_qr}
\end{figure*}

The background-subtracted distributions of the CFT output for the six signal decays considered are shown in \cref{fig:perf_qr}. \Cref{tab:tagging-performance} shows the corresponding tagging performance. Since the CFT needs only one loosely selected charged particle in the ROE as input, the tagging efficiency is almost 100\% and it is independent of the charmed hadron signal and of its decay mode. The mistag rate is independent of the signal decay mode, but it depends on the charmed hadron, given that different tagging categories contribute with different proportions depending on the charmed hadron signal (see, \eg, \cref{fig:tagger-split-by-categories}). The mistag rate being about $8\%$ larger in \Dp than \Dz decays can be attributed to the absence of the same-side soft-pion tagging category for \Dp decays. Similarly the increase in mistag rate for \Lc decays can be attributed to more important contributions from the proton tagging category for signal baryons compared to signal mesons. The mistag-rate difference between charm and anticharm signals is consistent with zero for neutral \D mesons. For charged hadrons the ROE must also be charged, and \dmistag shows significant deviations from zero due to the presence of detection asymmetries in the reconstruction of the ROE particles.

\begin{table*}[ht]
\centering
\begin{tabular}{cccccc}
\toprule
Signal decay &  \efftag (\%) & \defftag  (\%) & \mistag  (\%) & \dmistag  (\%) & \efftageff  (\%)\\
\midrule
$D^0\to K^-\pi^+$ & $99.974\pm0.004$ & $-0.002\pm0.007\phantom{-}$ & $19.09\pm0.08$ & $0.36\pm0.17$ & $38.22\pm 0.20$ \\ 
$D^0\to K^-\pi^+\pi^-\pi^+$ & $99.794\pm0.020$ & $0.042\pm0.039$ & $19.13\pm0.16$ & $0.40\pm0.32$ & $38.05\pm0.38$ \\ 
$D^0\to K^-\pi^+\pi^0$ & $99.967\pm0.006$ & $-0.006\pm0.012\phantom{-}$ & $19.34\pm0.13$ & $-0.22\pm0.26\phantom{-}$ & $37.58\pm0.32$ \\
$D^+\to K^-\pi^+\pi^+$ & $99.843\pm0.007$ & $-0.026\pm0.014\phantom{-}$ & $27.86\pm0.08$ & $0.80\pm0.16$ & $19.57\pm0.14$ \\ 
$D^+\to\KS\pi^+$ & $99.846\pm0.019$ & $0.037\pm0.038$ & $27.92\pm0.23$ & $1.83\pm0.46$ & $19.47\pm0.41$ \\ 
$\Lc \to pK^-\pi^+$ & $99.832\pm0.008$ & $-0.022\pm0.016\phantom{-}$ & $32.44\pm0.09$ & $0.52\pm0.18$ & $12.31\pm0.13$ \\ 
\bottomrule
\end{tabular}%
\caption{Performance of the CFT as determined on data in terms of tagging efficiency (\efftag), difference in tagging efficiency between charm and anticharm decays (\defftag), mistag rate (\mistag), difference in mistag rate between charm and anticharm decays (\dmistag), and average tagging power (\efftageff), for different signal decays. The uncertainties are statistical only.\label{tab:tagging-performance}}
\end{table*}

The results obtained with the $\Dz\to\Km\pip$ channel are
\begin{align}
\efftag &= (\phantom{-}99.974\pm 0.004\stat\pm 0.011\syst)\%\,,\\
\defftag &= (-\phantom{0}0.002\pm 0.007\stat\pm 0.004\syst)\%\,,\\
\mistag &= (\phantom{-}19.09\phantom{0}\pm 0.08\phantom{0}\stat\pm 0.17\phantom{0}\syst)\%\,,\\
\dmistag &= (\phantom{-0}0.36\phantom{0}\pm 0.17\phantom{0}\stat\pm 0.01\phantom{0}\syst)\%\,,\\
\efftageff &= (\phantom{-}38.22\phantom{0}\pm 0.20\phantom{0}\stat\pm 0.44\phantom{0}\syst)\%\,.\label{eq:result-average}
\end{align}
The systematic uncertainties are due to the background subtraction based on the \sPlot method~\cite{Pivk:2004ty}. They are computed using simulation by comparing the results obtained from the \sPlot method with those derived using generator-level information. Other sources of systematic uncertainties, such as contributions to the mistag rate due to the subtraction of the wrong-sign contribution, are negligible. As a consistency check, we evaluate the performance of the CFT by splitting the $\Dz\to\Km\pip$ signal sample in disjoint data-taking periods. No dependence on the data taking conditions is observed. The maximum number of input ROE tracks is varied from six to four and eight and the algorithm is retrained. We see a small degradation in performance when the number of ROE tracks is reduced and no substantial gain when it is increased.

\subsection{Calibration of the predicted dilution}
To fully exploit the tagging information in measurements of mixing and \CP asymmetries, the per-candidate dilution $r$ predicted by the CFT should be used in place of the average dilution resulting from the mistag fraction of \cref{sec:perf_res}.

\begin{figure}[ht]
\centering
\includegraphics[width=0.45\textwidth]{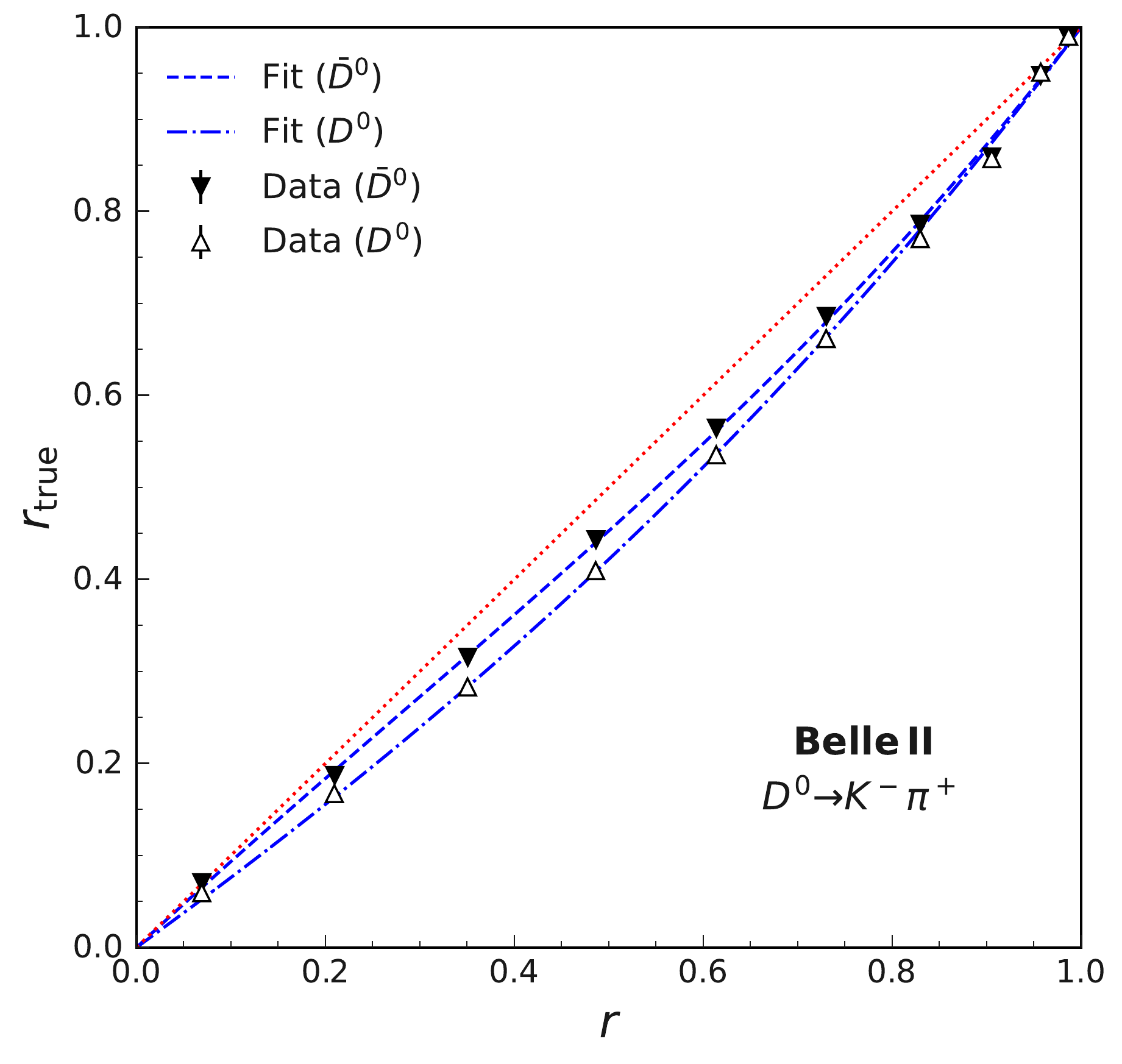}
\caption{True dilution as a function of the predicted dilution for $\Dz\to\Km\pip$ and $\Dzb\to\Kp\pim$ decays in data with fit projections overlaid. The bisector of the plane (red dotted line) represents the expected relation for perfectly calibrated predicted dilution.}\label{fig:calibration}
\end{figure}

To avoid biases, the per-candidate dilution is calibrated in data using the true dilution $r_{\rm true}$, computed starting from the true mistag fraction measured in the self-tagged $\Dz\to\Km\pip$ decays.  To maximally exploit the sample size, during the calibration, we consider all available $\Dz\to\Km\pip$ decays and do not discard 90\% of them as in \cref{sec:perf_res}. \Cref{fig:calibration} shows that the true dilution has a dependence on the predicted dilution that deviates from the expected linear $r_{\rm true}=r$ behaviour. The deviation occurs mostly for predicted dilution values dominated by opposite-side tagging categories (see \cref{fig:tagger-split-by-categories}), indicating that the algorithm is overconfidently assigning the flavor for this class of events. We fit a third-degree polynomial to the observed dependence. The polynomial is parametrized as
\begin{multline}\label{eq:calibration}
r_{\rm true}(r|q) = \left(p_1+q\frac{\Delta p_1}{2}\right) + \left(p_2+q\frac{\Delta p_2}{2}\right) \left(r-\frac{1}{2}\right) \\+ \left[2-4\left(p_1+q\frac{\Delta p_1}{2}\right)\right] \left(r-\frac{1}{2}\right)^2 + \left[4-4\left(p_2+q\frac{\Delta p_2}{2}\right)\right] \left(r-\frac{1}{2}\right)^3 \,,
\end{multline}
with the parameters $p_i$ accounting for the flavor-averaged calibration and the parameters $\Delta p_i$ for the differences observed between \Dz and \Dzb signal decays. To avoid exceeding the physical boundaries, \cref{eq:calibration} implies that $r_{\rm true}=r$ at the boundary values of $r=0$ and $r=1$. The results of the fit (\cref{tab:calibration}) are used to correct the CFT response and obtain a calibrated predicted per-candidate dilution $r_c$. Systematic uncertainties due to the background-subtraction procedure are included.

\begin{table*}[ht]
\centering
\begin{tabular}{cccccccc}
\toprule
 & & \multicolumn{6}{c}{Correlations (\%)}\\
 & & \multicolumn{3}{c}{stat.} & \multicolumn{3}{c}{syst.}\\
Coefficient & Value & $p_2$ & $\Delta p_1$ & $\Delta p_2$ & $p_2$ & $\Delta p_1$ & $\Delta p_2$\\
\midrule
$p_1$ & $0.437\pm0.001\pm0.007$ & $-44.6\phantom{-}$ & $3.3$ & $-0.7\phantom{-}$ & $-33.7\phantom{-}$ & $1.5$ & $5.3$\\
$p_2$ & $0.949\pm0.002\pm0.028$ & &  $-0.7\phantom{-}$ & $2.4$ & &  $2.4$ & $3.0$\\
$\Delta p_1$ & $-0.031\pm0.004\pm0.000\phantom{-}$ & & & $-44.6\phantom{-}$ & & & $-25.5\phantom{-}$\\
$\Delta p_2$ & $0.044\pm0.008\pm0.001$ & & &\\
\bottomrule
\end{tabular}%
\caption{Results of the fit to the true dilution as a function of the predicted dilution for $\Dz\to\Km\pip$ decays in data. The first uncertainties are statistical, the second systematic.}\label{tab:calibration}
\end{table*}

\begin{figure}
\centering
\includegraphics[width=0.45\textwidth]{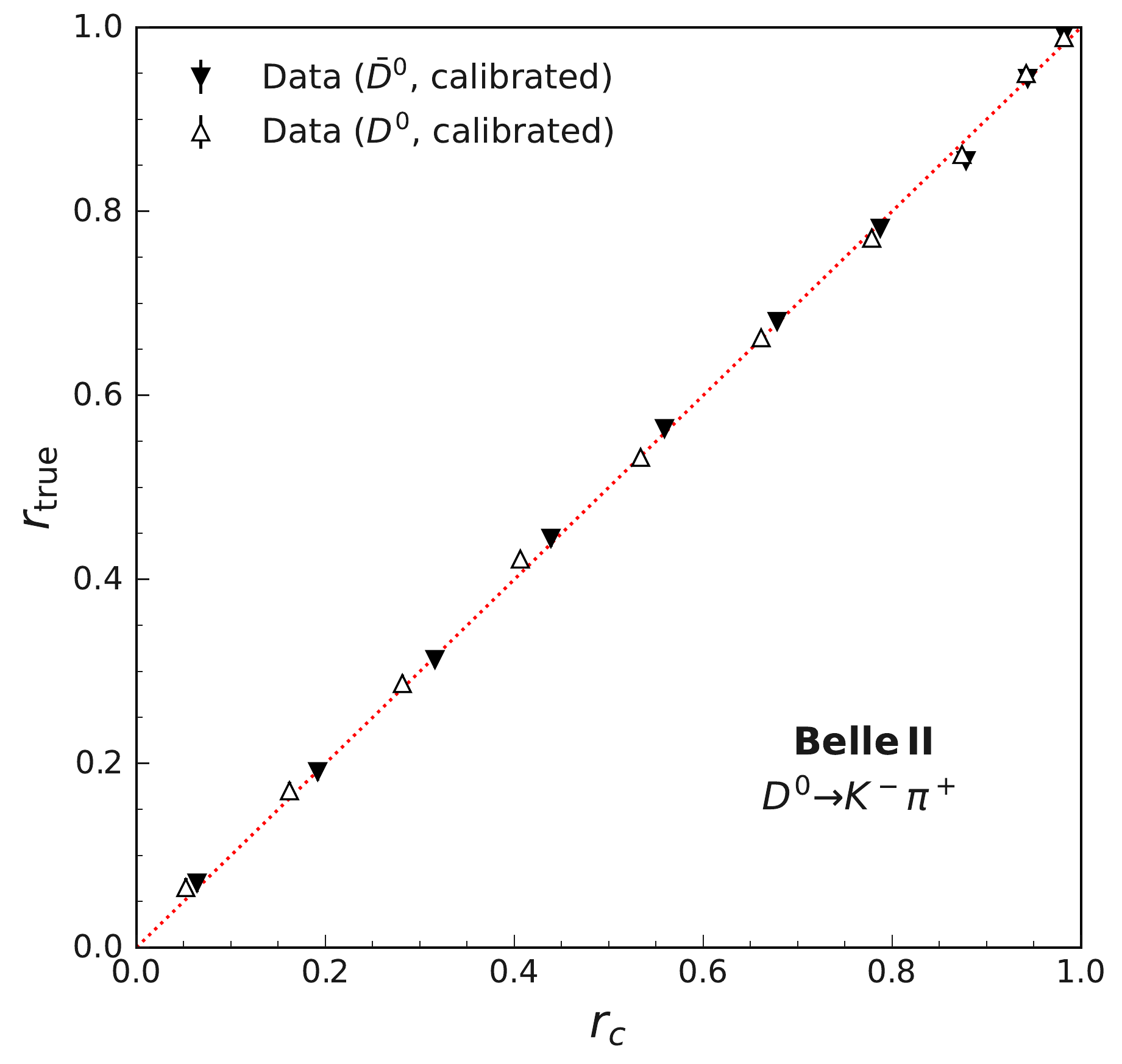}\\
\includegraphics[width=0.45\textwidth]{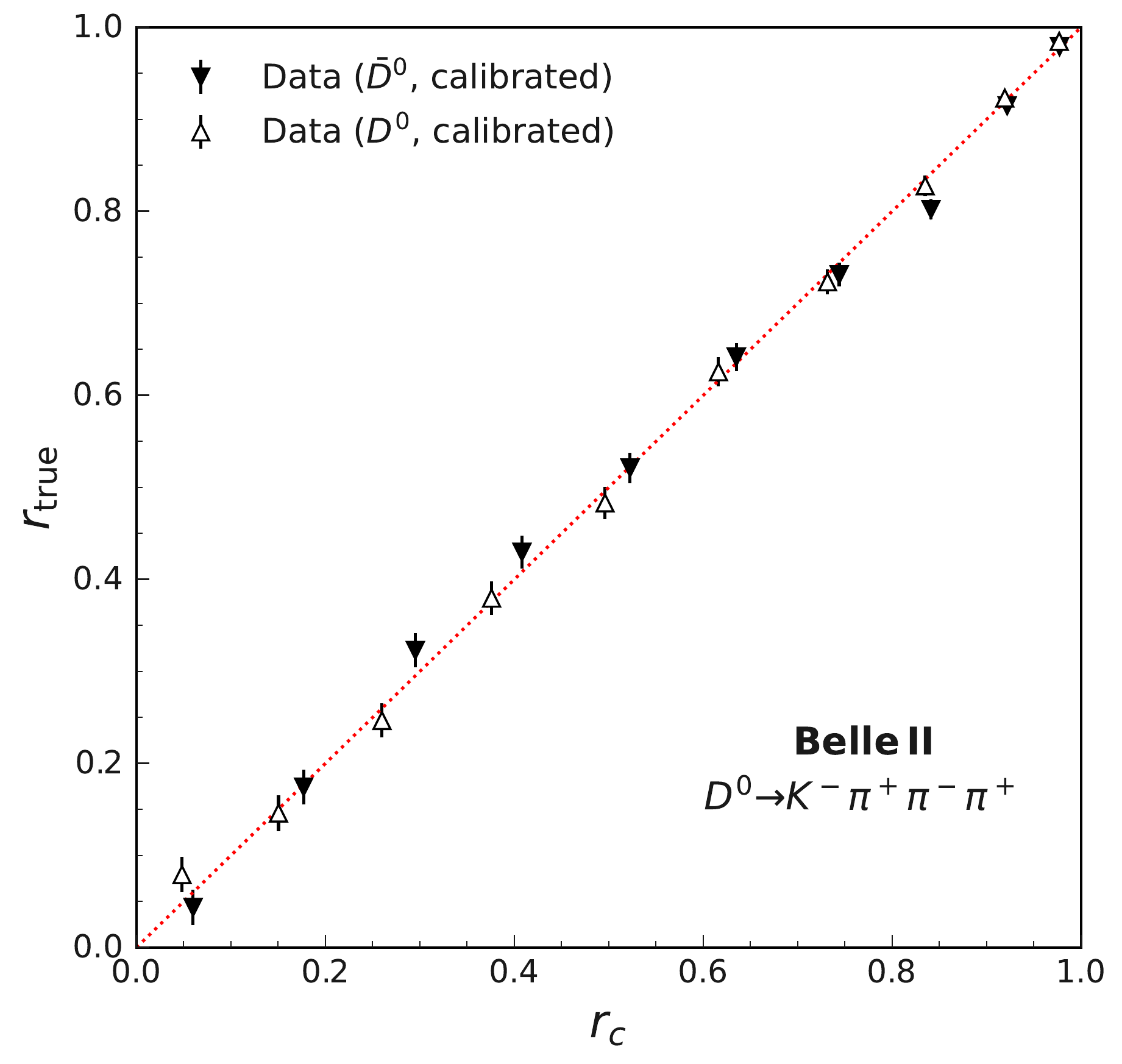}\\
\includegraphics[width=0.45\textwidth]{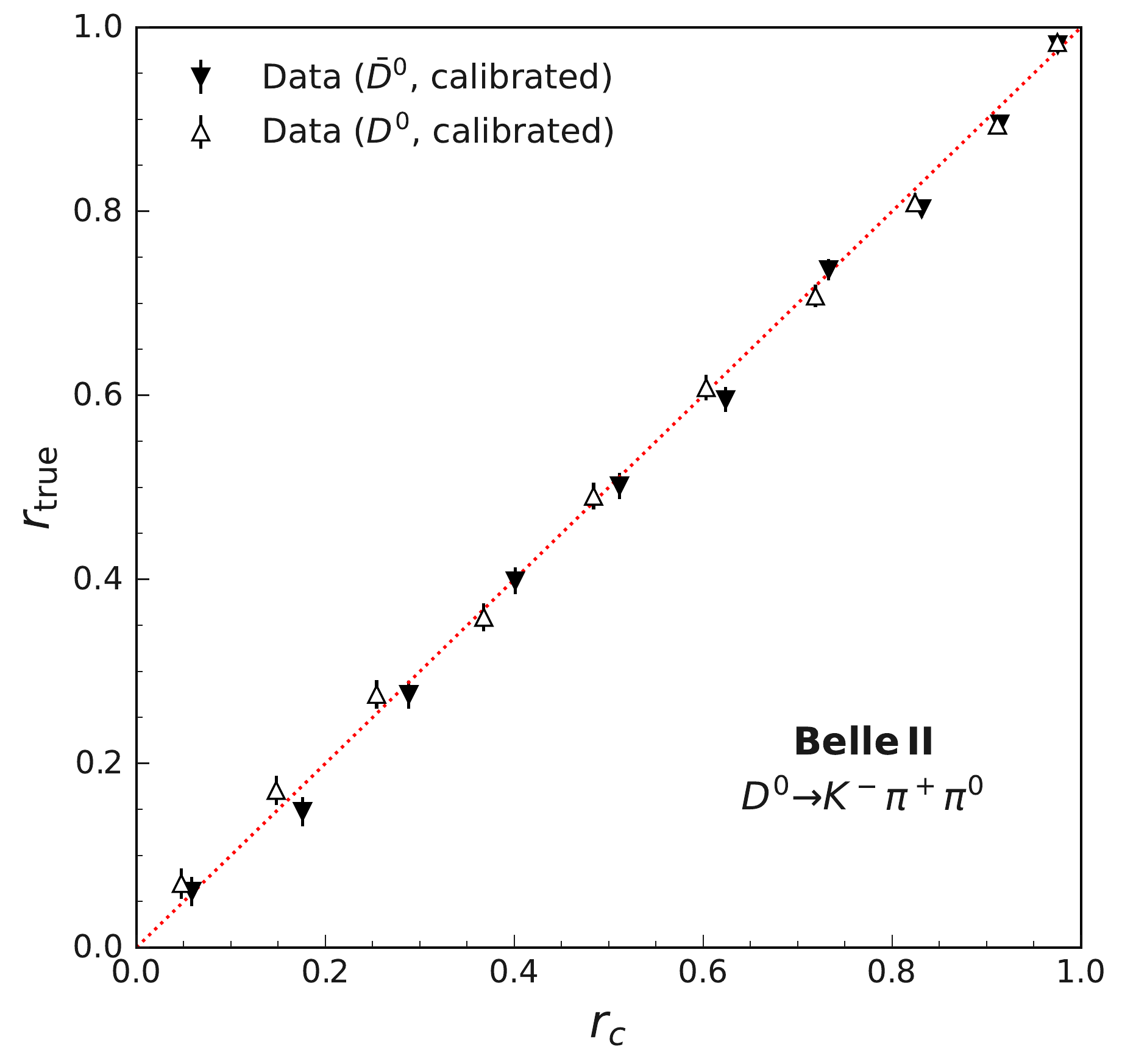}\\
\caption{True dilution as a function of the calibrated dilution for (top) $\Dz\to\Km\pip$, (middle) $\Dz\to\Km\pip\pip\pim$, and (bottom) $\Dz\to\Km\pip\piz$ decays in data.}\label{fig:calibrated-dilution-check}
\end{figure}

\Cref{fig:calibrated-dilution-check} shows the impact of the calibration on $\Dz\to\Km\pip$, $\Dz\to\Km\pip\pip\pim$, and $\Dz\to\Km\pip\piz$ signal decays. The calibrated dilution shows good agreement with the true dilution for all three decay modes indicating that the calibration obtained using $\Dz\to\Km\pip$ can be used for other signal decays.

With the calibrated per-candidate dilution, the tagging power is calculated, also for non flavor-specific decays, as 
\begin{equation}
\efftageff = \efftag\mean{r_c^2} = \frac{\sum_i w_ir_{c,\,i}^2}{\sum_i w_i}\,,
\end{equation}
where $i$ runs over the sample and $w_i$ is the \sPlot weight used to subtract the background. Including the systematic uncertainty due to the background subtraction with the \sPlot method, the tagging power for $\Dz\to\Km\pip$ decays is measured to be
\begin{equation}\label{eq:result-per-candidate}
\efftageff = \result\,.
\end{equation}
Since it fully exploits the information provided by the CFT, the tagging power based on the per-candidate dilution of \cref{eq:result-per-candidate} exceeds the tagging power based on the average dilution of \cref{eq:result-average}. 
 %
\section{Impact on physics\label{sec:impact}}
We estimate the effective increase in sample size in a typical mixing or \CP-asymmetry measurement that would otherwise rely exclusively on \Dstarp-tagged \Dz decays. We reconstruct a sample of $\Dz\to\Km\pip$ decays using \belletwo data corresponding to an integrated luminosity of 54.4\invfb. The sample is selected with the criteria of \cref{sec:reconstruction} and split into two disjoint subsets: events that are \Dstarp tagged, by explicitly reconstructing a $\Dstarp\to\Dz\pip$ decay and requesting the difference between \Dstarp and \Dz masses to satisfy $0.143<\Delta M<0.148\gevcc$; and events that are not \Dstarp tagged. 

The signal yields in the \Dstarp-tagged and non-\Dstarp-tagged samples are $125600\pm350$ and $388490\pm620$, respectively. The performance of the CFT on \Dstarp-tagged events is close to ideal. The subpercent mistag fraction is consistent with the level of non-\Dstarp background candidates made of \Dz signal decays associated with unrelated soft pions. The tagging power on non-\Dstarp-tagged events, computed using the calibrated per-candidate dilution, is $(32.71\pm0.05\stat)\%$. By multiplying the signal yield and the tagging power in such a configuration, we estimate that the CFT provides an additional $127080\pm280$ tagged signal decays for mixing and \CP-asymmetry measurements, effectively doubling the sample size compared to \Dstarp-tagged events. However, such an increase in sample size compared to \Dstarp-tagged decays is accompanied by an increased background. Hence, doubling the sample size is not expected to correspond to a factor $\sqrt{2}$ increase in the precision of the measurement.

\begin{figure*}[t]
\centering
\includegraphics[width=0.49\textwidth]{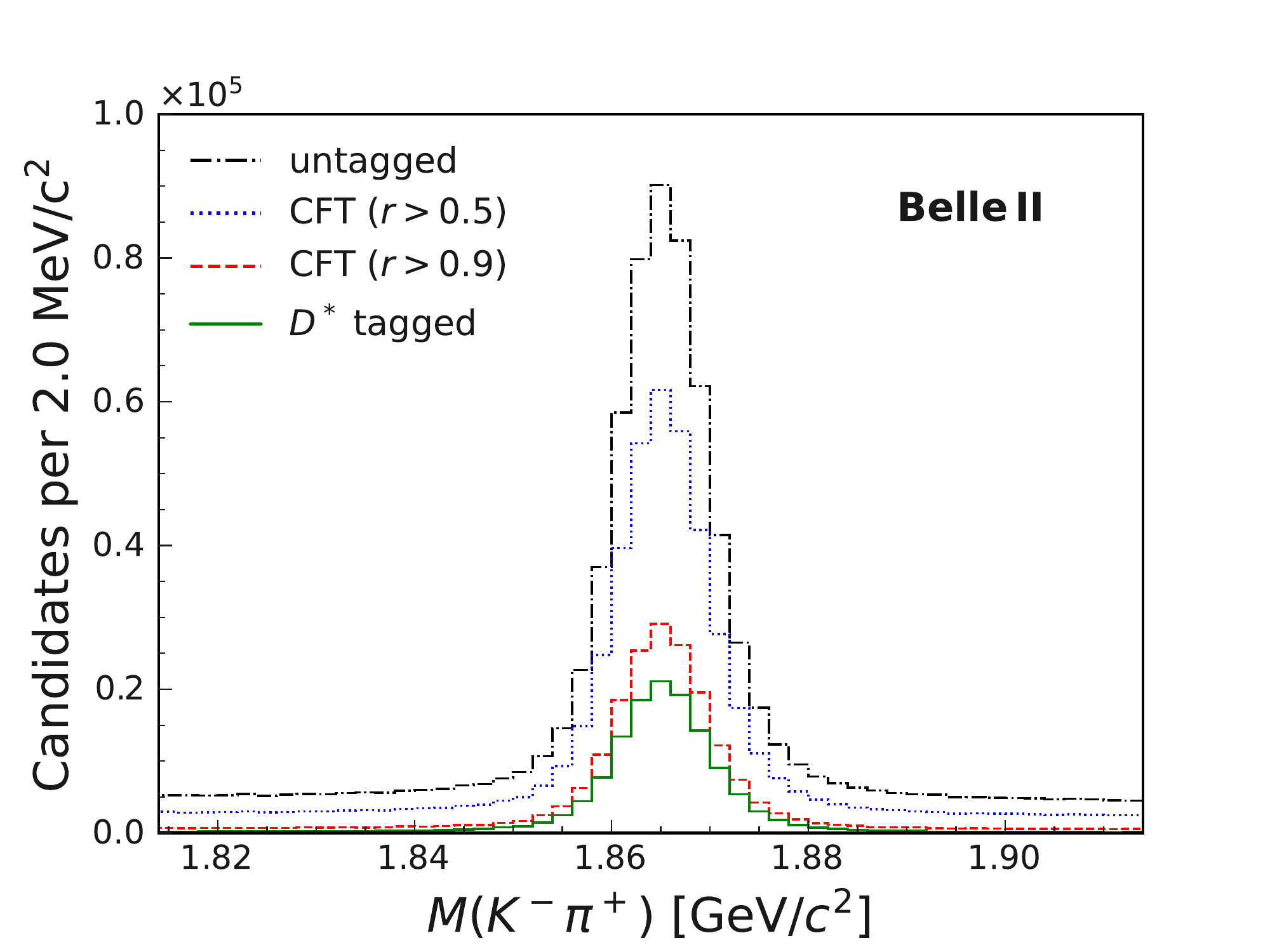}\hfil
\includegraphics[width=0.49\textwidth]{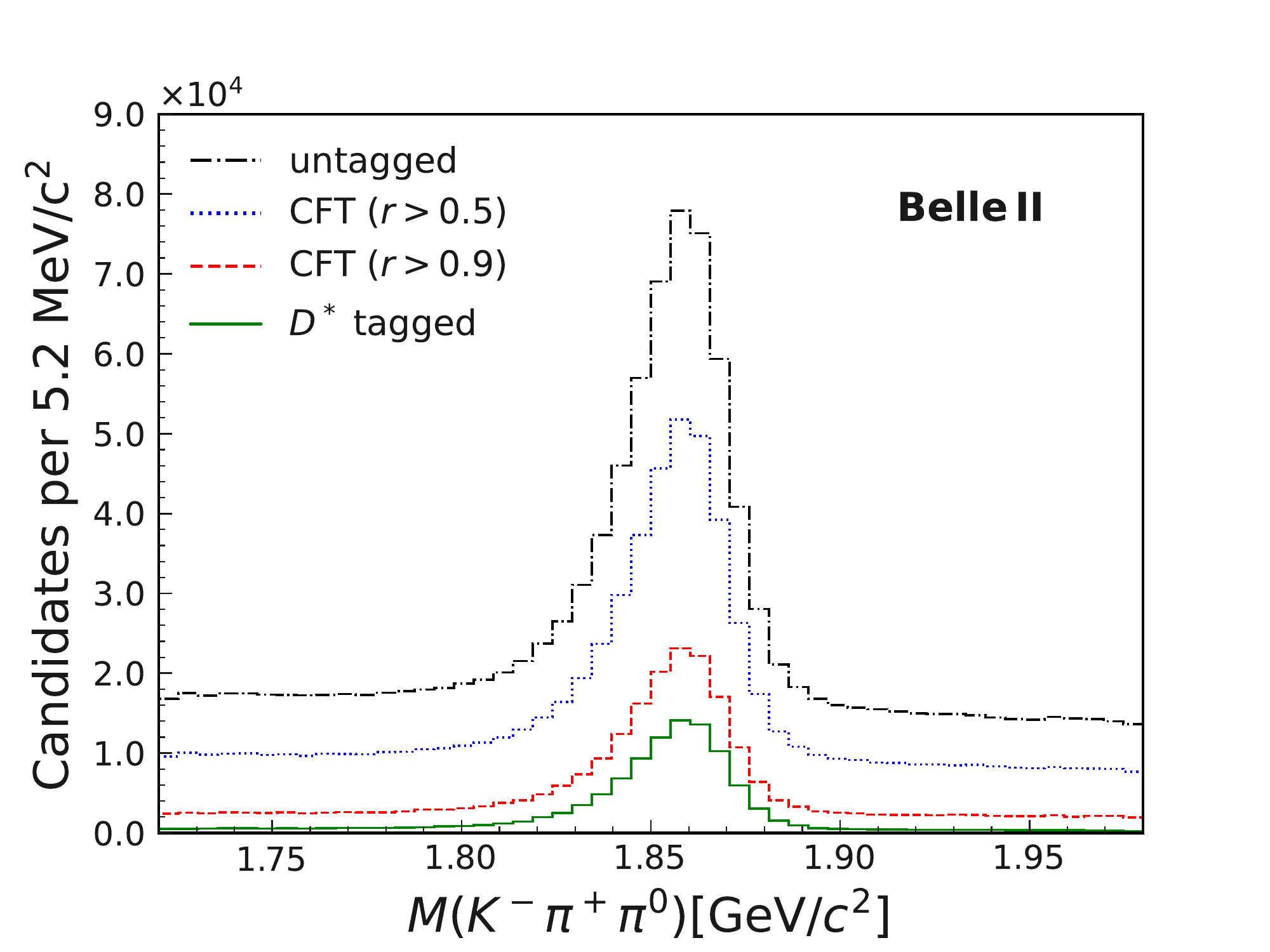}\\
\caption{Mass distributions for (left) $\Dz\to\Km\pip$ decays and (right) $\Dz\to\Km\pip\piz$ decays reconstructed in data with different requirements on the predicted (uncalibrated) dilution in comparison with \Dstarp-tagged decays. For the selections shown the $\Dz\to\Km\pip$ signal 
purities are 0.94 (\Dstarp-tagged), 0.84 (CFT, $r>0.9$), 0.73 (CFT, $r>0.5$), and 0.67 (untagged). For $\Dz\to\Km\pip\piz$ decays the signal purities are 0.80 (\Dstarp-tagged), 0.53 (CFT, $r>0.9$), 0.38 (CFT, $r>0.5$), and 0.34 (untagged).}
\label{fig:impt_compare_mass}
\end{figure*}

\begin{figure}[t]
\centering
\includegraphics[width=0.49\textwidth]{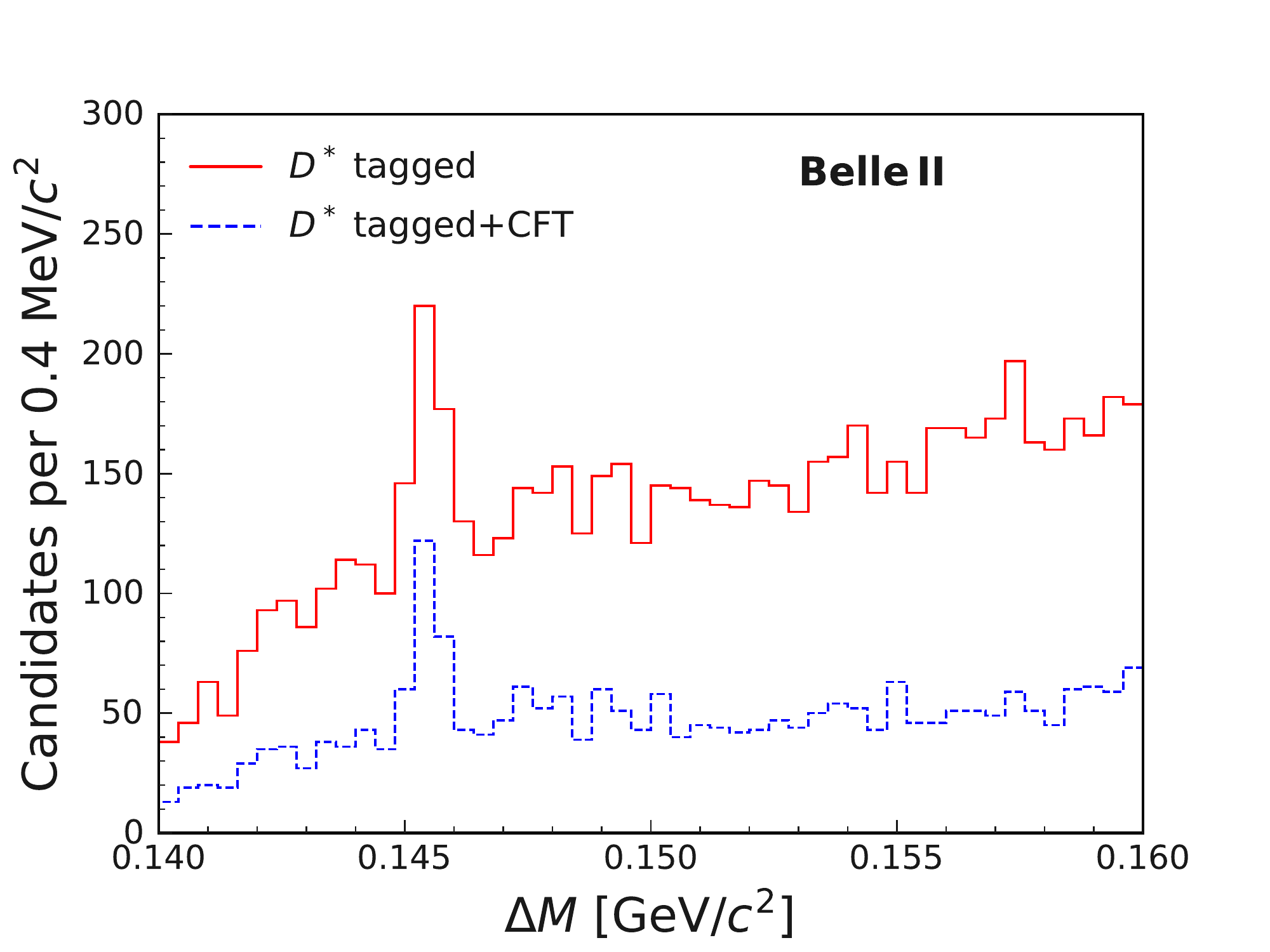}
\caption{Distribution of the difference between \Dstarp and \Dz masses for wrong-sign $\Dstarp\to\Dz(\to\Kp\pim\piz)\pip$ decays reconstructed in data and selected with and without the requirement $q_{\pis}q>0$.}
\label{fig:impt_ws}
\end{figure}

In addition, the CFT output distribution is expected to provide some discrimination between signal and background. Such separation can be effectively used in a fit that has the calibrated per-candidate dilution as an observable or, as shown in \cref{fig:impt_compare_mass}, it can be used as part of the selection requirements to improve the signal purity. Such a feature may be particularly valuable for analyses that do not require tagging but reconstruct charmed hadrons with small signal-to-background ratios. An example is shown in \cref{fig:impt_ws} for wrong-sign $\Dstarp\to\Dz(\to\Kp\pim\piz)\pip$ decays selected in a sample of \belletwo data corresponding to 54.4\invfb. The CFT is used in the sample selection to confirm the tag provided by the \Dstarp decay with the requirement $q_{\pis}q>0$, where $q_{\pis}$ is the soft-pion charge. With only a 24\% loss of signal yield, the signal-to-background ratio in the resulting \textit{doubly tagged} sample is roughly doubled compared to the sample where the CFT is not used.

 %
\section{Summary}%
We developed a novel charm-flavor tagging algorithm for \belletwo that determines the production flavor of a signal neutral \D meson. The algorithm exploits the correlation between the production flavor and the electric charges of particles reconstructed in the rest of the event, \ie, those originating from the decay of the other charmed hadron produced in the $\epem\to\cquark\cquarkbar$ process and those produced in association with the signal \D meson (\eg, in the decay of a parent \Dstarpm meson). The tagger uses boosted decision trees trained on simulated data. Its response is calibrated and evaluated using several self-tagged decays of charmed hadrons reconstructed in \lumi of \belletwo data. The effective tagging efficiency is measured in data to be \result, independent of the signal neutral-\D decay mode. This new tagger will roughly double the effective sample size for \CP-violation and charm-mixing measurements that so far have relied exclusively on neutral \D mesons originating from \Dstarpm decays. Moreover, the tagger can be effectively used to suppress backgrounds for measurements in which tagging is not required, making it a more general tool for analyses of charmed hadrons at \belletwo. While developed explicitly for \belletwo, the basic principles of this new tagger are adequate for other experiments, including those at hadron colliders where charmed hadrons are predominantly produced from $\cquark\cquarkbar$ pairs. 
 
%
%
This work, based on data collected using the Belle II detector, which was built and commissioned prior to March 2019, was supported by
Science Committee of the Republic of Armenia Grant No.~20TTCG-1C010;
Australian Research Council and research Grants
No.~DE220100462,
No.~DP180102629,
No.~DP170102389,
No.~DP170102204,
No.~DP150103061,
No.~FT130100303,
No.~FT130100018,
and
No.~FT120100745;
Austrian Federal Ministry of Education, Science and Research,
Austrian Science Fund
No.~P~31361-N36
and
No.~J4625-N,
and
Horizon 2020 ERC Starting Grant No.~947006 ``InterLeptons'';
Natural Sciences and Engineering Research Council of Canada, Compute Canada and CANARIE;
Chinese Academy of Sciences and research Grant No.~QYZDJ-SSW-SLH011,
National Natural Science Foundation of China and research Grants
No.~11521505,
No.~11575017,
No.~11675166,
No.~11761141009,
No.~11705209,
and
No.~11975076,
LiaoNing Revitalization Talents Program under Contract No.~XLYC1807135,
Shanghai Pujiang Program under Grant No.~18PJ1401000,
Shandong Provincial Natural Science Foundation Project~ZR2022JQ02,
and the CAS Center for Excellence in Particle Physics (CCEPP);
the Ministry of Education, Youth, and Sports of the Czech Republic under Contract No.~LTT17020 and
Charles University Grant No.~SVV 260448 and
the Czech Science Foundation Grant No.~22-18469S;
European Research Council, Seventh Framework PIEF-GA-2013-622527,
Horizon 2020 ERC-Advanced Grants No.~267104 and No.~884719,
Horizon 2020 ERC-Consolidator Grant No.~819127,
Horizon 2020 Marie Sklodowska-Curie Grant Agreement No.~700525 "NIOBE"
and
No.~101026516,
and
Horizon 2020 Marie Sklodowska-Curie RISE project JENNIFER2 Grant Agreement No.~822070 (European grants);
L'Institut National de Physique Nucl\'{e}aire et de Physique des Particules (IN2P3) du CNRS (France);
BMBF, DFG, HGF, MPG, and AvH Foundation (Germany);
Department of Atomic Energy under Project Identification No.~RTI 4002 and Department of Science and Technology (India);
Israel Science Foundation Grant No.~2476/17,
U.S.-Israel Binational Science Foundation Grant No.~2016113, and
Israel Ministry of Science Grant No.~3-16543;
Istituto Nazionale di Fisica Nucleare and the research grants BELLE2;
Japan Society for the Promotion of Science, Grant-in-Aid for Scientific Research Grants
No.~16H03968,
No.~16H03993,
No.~16H06492,
No.~16K05323,
No.~17H01133,
No.~17H05405,
No.~18K03621,
No.~18H03710,
No.~18H05226,
No.~19H00682, %
No.~22H00144,
No.~26220706,
and
No.~26400255,
the National Institute of Informatics, and Science Information NETwork 5 (SINET5), 
and
the Ministry of Education, Culture, Sports, Science, and Technology (MEXT) of Japan;  
National Research Foundation (NRF) of Korea Grants
No.~2016R1\-D1A1B\-02012900,
No.~2018R1\-A2B\-3003643,
No.~2018R1\-A6A1A\-06024970,
No.~2018R1\-D1A1B\-07047294,
No.~2019R1\-I1A3A\-01058933,
No.~2022R1\-A2C\-1003993,
and
No.~RS-2022-00197659,
Radiation Science Research Institute,
Foreign Large-size Research Facility Application Supporting project,
the Global Science Experimental Data Hub Center of the Korea Institute of Science and Technology Information
and
KREONET/GLORIAD;
Universiti Malaya RU grant, Akademi Sains Malaysia, and Ministry of Education Malaysia;
Frontiers of Science Program Contracts
No.~FOINS-296,
No.~CB-221329,
No.~CB-236394,
No.~CB-254409,
and
No.~CB-180023, and No.~SEP-CINVESTAV research Grant No.~237 (Mexico);
the Polish Ministry of Science and Higher Education and the National Science Center;
the Ministry of Science and Higher Education of the Russian Federation,
Agreement No.~14.W03.31.0026, and
the HSE University Basic Research Program, Moscow;
University of Tabuk research Grants
No.~S-0256-1438 and No.~S-0280-1439 (Saudi Arabia);
Slovenian Research Agency and research Grants
No.~J1-9124
and
No.~P1-0135;
Agencia Estatal de Investigacion, Spain
Grant No.~RYC2020-029875-I
and
Generalitat Valenciana, Spain
Grant No.~CIDEGENT/2018/020
Ministry of Science and Technology and research Grants
No.~MOST106-2112-M-002-005-MY3
and
No.~MOST107-2119-M-002-035-MY3,
and the Ministry of Education (Taiwan);
Thailand Center of Excellence in Physics;
TUBITAK ULAKBIM (Turkey);
National Research Foundation of Ukraine, project No.~2020.02/0257,
and
Ministry of Education and Science of Ukraine;
the U.S. National Science Foundation and research Grants
No.~PHY-1913789 %
and
No.~PHY-2111604, %
and the U.S. Department of Energy and research Awards
No.~DE-AC06-76RLO1830, %
No.~DE-SC0007983, %
No.~DE-SC0009824, %
No.~DE-SC0009973, %
No.~DE-SC0010007, %
No.~DE-SC0010073, %
No.~DE-SC0010118, %
No.~DE-SC0010504, %
No.~DE-SC0011784, %
No.~DE-SC0012704, %
No.~DE-SC0019230, %
No.~DE-SC0021274, %
No.~DE-SC0022350; %
and
the Vietnam Academy of Science and Technology (VAST) under Grant No.~DL0000.05/21-23.

These acknowledgements are not to be interpreted as an endorsement of any statement made
by any of our institutes, funding agencies, governments, or their representatives.

We thank the SuperKEKB team for delivering high-luminosity collisions;
the KEK cryogenics group for the efficient operation of the detector solenoid magnet;
the KEK computer group and the NII for on-site computing support and SINET6 network support;
and the raw-data centers at BNL, DESY, GridKa, IN2P3, INFN, and the University of Victoria for offsite computing support.
 
\bibliographystyle{belle2}
\providecommand{\href}[2]{#2}\begingroup\raggedright\endgroup

\end{document}